\documentclass{jfm_draft}

\usepackage[utf8]{inputenc}
\usepackage{graphicx}
\usepackage{overpic}
\usepackage[sort,numbers]{natbib}
\usepackage{mathtools}
\usepackage{amssymb}
\usepackage{amsbsy}
\usepackage{bm}
\usepackage{color}
\usepackage{tikz}
\usepackage[cal=euler]{mathalpha}
\usepackage{xcolor} 
\usepackage[hidelinks]{hyperref}
\usepackage[nameinlink,noabbrev]{cleveref}
\usepackage{siunitx}

\usepackage{my_macros}

\allowdisplaybreaks

\title{Generalised Jeffery's equations for rapidly spinning particles. Part 2: Helicoidal objects with chirality}
\date{\today}

\shorttitle{Part 2: Generalised Jeffery's equations for fast-spinning helicoidal objects}
\shortauthor{Dalwadi, Moreau, Gaffney, Walker, and Ishimoto}
\author{M. P. Dalwadi\aff{1\corresp{\email{m.dalwadi@ucl.ac.uk}}}, C. Moreau\aff{2}, E. A. Gaffney\aff{3}, B. J. Walker\aff{1,4}, K. Ishimoto\aff{2}}

\affiliation{\aff{1}Department of Mathematics, University College London, London, WC1H 0AY, UK
\aff{2}Research Institute for Mathematical Sciences, Kyoto University, Kyoto, 606-8502, Japan
\aff{3}Wolfson Centre for Mathematical Biology, Mathematical Institute, University of Oxford, Oxford, OX2 6GG, UK
\aff{4}Department of Mathematical Sciences, University of Bath, Bath, BA2 7AY, UK}

\date{}

\begin{document}

\maketitle

\section*{Abstract}
In this two-part study, we investigate the motion of rigid, active objects in shear Stokes flow, focusing on bodies that induce rapid rotation as part of their activity. In Part 2, we derive and analyse governing equations for rapidly spinning complex-shaped particles - general helicoidal objects with chirality. Using the multiscale framework we develop in Part 1 \citep{dalwadi2023emergentPart1}, we systematically derive emergent equations of motion for the angular and translational dynamics of these chiral spinning objects. We show that the emergent dynamics due to rapid rotation can be described by effective generalised Jeffery's equations, which differ from the classic versions via the inclusion of additional terms that account for chirality and other asymmetries. Furthermore, we use our analytic results to characterise and quantify the explicit effect of rotation on the effective hydrodynamic shape of the chiral objects, significantly expanding the scope of Jeffery's seminal study.

\section{Introduction}
\label{sec: Introduction}

The complex dynamics of objects in fluid flow are known to depend strongly on an object's shape, with the early study of \citet{Jeffery1922} explicitly capturing the behaviour of passive spheroidal particles in shear Stokes flow. Later extensions by \citet{Bretherton1962} and \citet{Brenner1964b} widen the range of passive objects to which \citeauthor{Jeffery1922}'s approach applies, with geometry playing a fundamental role in determining the dynamics.

In this two-part study, we consider the emergent dynamics of rigid \emph{active} objects. Inspired by the locomotion of flagellated bacterial swimmers \citep{marcos2012bacterial}, we consider swimmers whose activity consists of rapid rotation while propelling themselves through the surrounding fluid. In Part 1 \citep{dalwadi2023emergentPart1}, we developed a multiscale framework to analyse rapidly rotating particles in Stokes flow, and applied it to investigate spheroidal objects in shear flow, which follow Jeffery's equations. Here, in Part 2, we broaden our analysis to general helicoidal objects (described in detail below), including \emph{chiral} particles, whose passive dynamics are governed by generalised versions of Jeffery's equations \citep{Ishimoto2020b,Ishimoto2020a}. The dynamics of chiral bodies are generally more intricate than achiral bodies since chiral objects induce additional hydrodynamic interactions. The importance of chiral effects has been identified and utilised in theoretical and experimental studies across many different areas, including the drift-induced separation of chiral objects \citep{marcos2009separation,eichhorn2010microfluidic,aristov2013separation,ro2016chiral}, chirality-affected rheotaxis in bacterial and artificial swimmers \citep{mathijssen2019oscillatory,marcos2012bacterial,jing2020chirality,zheng2023swimming,zottl2022asymmetric,khatri2022diffusion}, the migration of chiral DNA-like objects \citep{chen2011dynamical}, and the preferential rotation of chiral dipoles \citep{kramel2016preferential}.

Certain geometric symmetries generate specific simplifications to the hydrodynamic resistance tensor associated with the object in Stokes flow. However, while the hydrodynamic resistance tensor depends strongly on an object's geometry, it does not uniquely define the shape. That is, there exist objects with the same simplified hydrodynamic resistance tensor but without the associated geometric symmetries. Sharing the same form of the hydrodynamic resistance tensor defines a \emph{hydrodynamic symmetry}. Importantly, this means there is a difference between the hydrodynamic symmetry of an object -- the properties of its dynamics in flow -- and its geometric features.

In Part 2 of this two-part study, we consider swimmers that possess helicoidal hydrodynamic symmetry about a swimmer-fixed axis, and we refer to objects that satisfy this type of symmetry as \emph{helicoidal objects}. This symmetry, introduced by \citet{Brenner1964a,Brenner1964b} and recounted recently by \citet{Ishimoto2020b}, generalises the geometric notion of rotational symmetry in the context of fluid mechanics. Specifically, helicoidal symmetry means that the hydrodynamic resistance tensor associated with the object is invariant under rotations by $\pm\heliangle$ about a swimmer-fixed axis for some fixed $\heliangle\not\in\{0,\pi,2\pi/3\}$ (with the excluded cases noted to be degenerate by \citet{Brenner1964b}).

The distinction between hydrodynamic and geometric symmetry is important because it is not straightforward to geometrically characterise the properties of an object with hydrodynamic symmetry. For example, objects that have $n$-fold rotational symmetry for some $n \geqslant 4$ are hydrodynamically helicoidal \citep{Brenner1964a,Ishimoto2020a}, but objects with geometric helical symmetry (e.g. a simple helix of finite length) are not helicoidal in general. Of particular note, while axisymmetric objects follow Jeffery's equations as stated by \citet{Bretherton1962}, not all objects governed by Jeffery's equations are axisymmetric. In light of this, we characterise the particles described by the analysis of Part 1 of this two-part study (i.e. those that follow Jeffery's equations) as `Jeffery bodies'. We emphasise that this definition includes simple spheroids.

The general active helicoidal objects we consider in this part are generalised versions of these Jeffery bodies. As identified in \citet{Ishimoto2020a}, the dynamics of a passive helicoidal particle in shear flow are governed by \emph{generalised} Jeffery orbits comprising six characteristic parameters, in contrast to only one for Jeffery bodies (the Bretherton parameter $\Breth$). When we introduce the governing dynamical system for active particles later, we discuss the role of these six parameters, along with subcases of interest and correspondences with geometric symmetries of the object. A detailed discussion of chirality, general helicoidal objects, and their associated contributions to the governing equations of motion for passive objects can be found in \citet{Ishimoto2020b}.

In our study, we specifically allow the axis of the self-propelled spinning to deviate from the axis of symmetry, as is the case for a wiggling bacterium \citep{hyon2012wiggling, thawani2018trajectory} and a wobbling magnetised helix \citep{man2013wobbling}.
In these contexts, the timescale of activity-driven spinning is typically much faster than that of reorientation by an imposed flow field. Motivated by these separated timescales, we analyse the dynamics using the asymptotic method of multiple scales \citep{hinch_1991,Bender1999}, as in Part 1 and several recent works \citep{Walker2022a,Gaffney2022,Ma2022b}. In particular, we derive effective governing equations for the emergent dynamics, systematically accounting for the complex nonlinear interaction between rapid rotation and the slower effects of the flow.

Hence, in this second part of our two-part study, we consider the dynamics of a three-dimensional, self-propelled chiral object with helicoidal symmetry, undergoing rapid spinning due to its own activity, and interacting with an externally imposed three-dimensional shear flow. In \S \ref{sec: Governing equations}, we present the general governing equations for the system, including additional terms not present in Part 1 that account for chirality and other asymmetries of the object. In \S \ref{sec: setup multiscle}, we set up the machinery for our multiple scales analysis then, in \S \ref{Sec: emergent angular dynamics} and \S \ref{sec: emergent trans dyn}, we perform the analysis for both rotation and translation, respectively, systematically deriving effective governing equations that explicitly capture the effects of rapid spinning on the overall dynamics. As one may expect, the effective dynamics we derive for general helicoidal particles in Part 2 are significantly richer than those we derive for simple spheroidal particles in Part 1. Hence, we summarise the key physical results and implications of the emergent dynamics we derive through our analysis in a non-technical manner in \S \ref{sec: results}. Finally, we conclude with a discussion of our study in \S \ref{sec: Discussion}.

\section{Governing equations}
\label{sec: Governing equations}

Our physical setup in Part 2 is similar to that in Part 1, but now with a more complex swimmer geometry. That is, we now consider a general helicoidal swimmer, as discussed in \S \ref{sec: Introduction}, in the presence of a far-field shear flow. This will result in additional hydrodynamic effects due to object chirality and other asymmetries. We scale time with inverse shear rate, and space with a characteristic swimmer length, working in dimensionless quantities henceforth. Specifically, we consider the motion of a rigid, self-propelled helicoidal object in a shear flow, which has a swimming velocity $\vel$ and angular velocity $\angvel$ in a quiescent fluid. As before, these propulsion and rotation vectors are fixed in direction and magnitude in a swimmer-fixed basis, but the orientation of this swimmer basis will vary rapidly in the laboratory frame through its dependence on $\angvel$.

\begin{figure}
    \centering
    \includegraphics[width=0.7\textwidth]{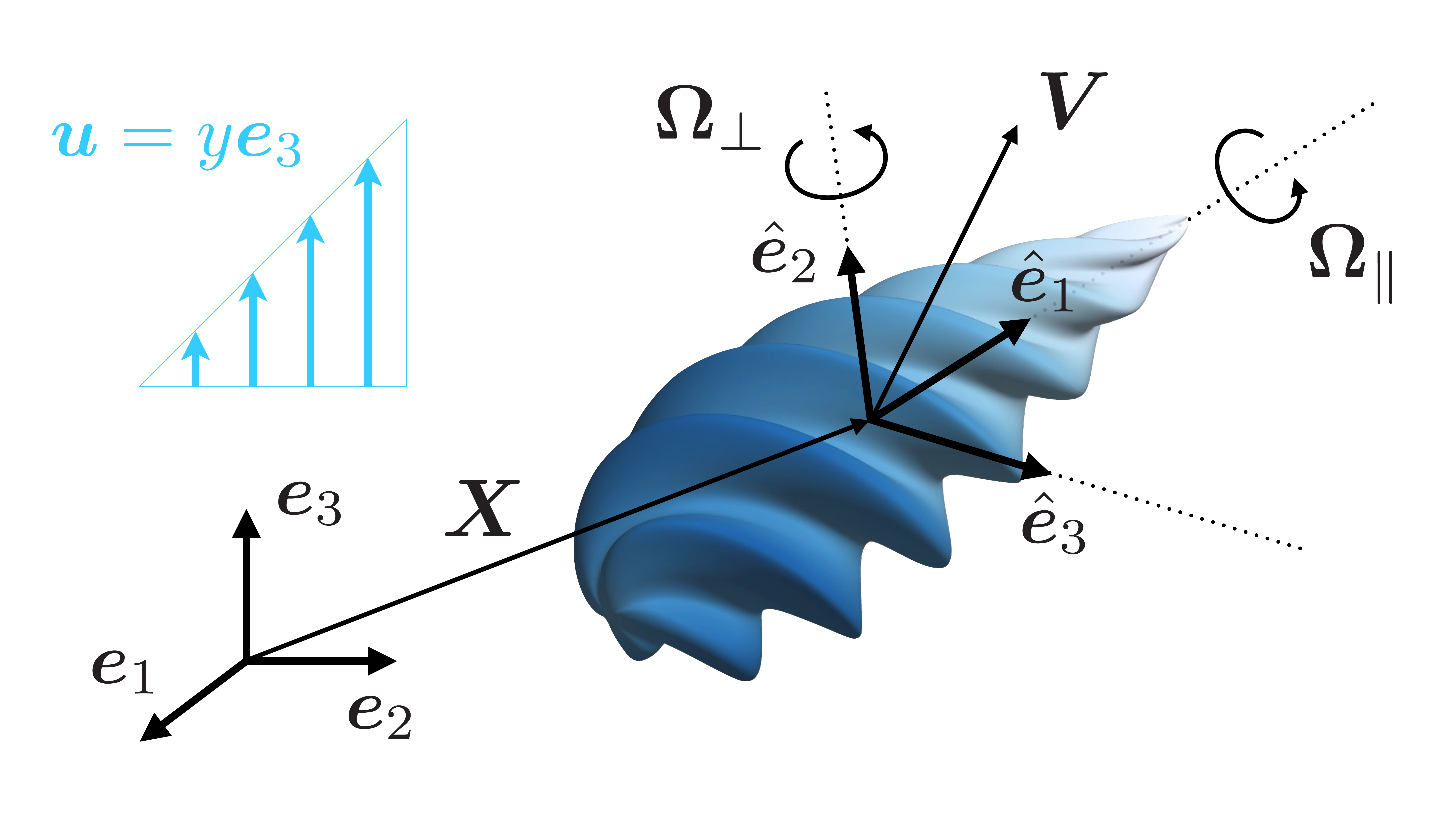}
    \vspace{-3mm}
    \caption{A schematic of the notation and the physical setup we consider in Part 2. We investigate the dynamics of a chiral, helicoidal swimmer with axis of symmetry $\ehat{1}$. The swimmer has self-generated translational and rotational velocities $\vel = \velscala\ehat{1} + \velscalb\ehat{2} + \velscalc\ehat{3}$ and $\angvel = \omb\ehat{1} + \oma\ehat{2}$, respectively, and these interact with a background shear flow $\flowvel = y \e{3}$.}
    \label{fig: setup}
\end{figure}

We define the swimmer-fixed axis of helicoidal symmetry by $\ehat{1}$. Therefore, we may take $\ehat{2}$ such that the self-generated angular velocity $\angvel$ is in a plane spanned by $\ehat{1}$ and $\ehat{2}$, where $\angvel$ makes an angle of $\angl$ with $\ehat{1}$. Therefore, we may write $\angvel = \omb\ehat{1} + \oma\ehat{2}$, with $\omb$ and $\oma$ being the constant components of angular velocity that are parallel and perpendicular, respectively, to the axis of helicoidal symmetry. This generates the relationship $\tan \angl = \oma/\omb$. We then define $\ehat{3} = \ehat{1} \times \ehat{2}$. In this swimmer-fixed basis, we write the self-generated propulsion $\vel = \velscala\ehat{1} + \velscalb\ehat{2} + \velscalc\ehat{3}$. The position of the particle is given by $\Xvecpos = \Xpos \e{1} + \Ypos \e{2} + \Zpos \e{3}$ with respect to the orthonormal basis $\{\e{1},\e{2},\e{3}\}$ of the laboratory frame. These vectors are illustrated in \Cref{fig: setup}.

Finally, we impose the far-field flow. Specifically, we are interested in the motion of the particle in the presence of a far-field shear Stokes flow with velocity field $\flowvel(x,y,z) = y \e{3}$, with coordinates $x,y,z$ in the laboratory frame. The flow interacts with the particle; we derive the resulting governing equations of motion for the particle in Appendix \ref{sec: derivation eom}, and present the resulting equations below. The dynamics for the orientation of the swimmer frame are given in terms of the Euler angles $(\theta, \psi, \phi)$, also defined formally in Appendix \ref{sec: derivation eom}.

The rotational dynamics are given by
\begin{subequations}
\label{eq: full gov eq}
\begin{align}
\label{eq: theta eq}
\dbyd{\theta}{\tstandard} &= \oma \cos \psi + \hb(\theta,\phi; \Breth,\Ish), \\
\label{eq: psi eq}
\dbyd{\psi}{\tstandard} &= \omb - \oma \dfrac{\cos \theta \sin \psi}{\sin \theta} + \hc(\theta,\phi; \Breth,\Ish,\DIsh), \\
\label{eq: phi eq}
\dbyd{\phi}{\tstandard} &= \oma \dfrac{\sin \psi}{\sin \theta} + \ha(\theta,\phi; \Breth,\Ish),
\end{align}
\end{subequations}
where the functions $h_i = f_i + g_i$ ($i = 1, 2, 3$) capture the effects of the Stokes flow interacting with the swimmer. The $f_i$ encode the rotational effects of the achiral aspects of the swimmer (and are the same as in Part 1). These functions are
\begin{subequations}
\label{eq: f functions}
\begin{align}
\fb(\theta,\phi; \Breth) &:= -\dfrac{ \Breth}{2} \cos \theta \sin \theta \sin 2 \phi, \\
\fc(\theta,\phi; \Breth) &:= \dfrac{\Breth}{2} \cos \theta \cos 2 \phi, \\
\fa(\phi; \Breth) &:= \dfrac{1}{2} \left(1 - \Breth \cos 2 \phi\right),
\end{align}
\end{subequations}
where $\Breth$ is the shape-capturing Bretherton parameter \citep{Bretherton1962}, which typically satisfies $\abs{\Breth}<1$ for all but the most elongated of bodies \citep{Bretherton1962, singh2013rigid}.

The $g_i$ encode the rotational effects of the chiral aspects of the swimmer, and were therefore not present in the spheroidal analysis of Part 1. These chiral functions are
\begin{subequations}
\label{eq: g functions}
\begin{align}
\label{eq: g functions 1}
\gb(\theta,\phi; \Ish) &:=  -\dfrac{\Ish}{2} \sin \theta \cos 2 \phi, \\
\gc(\theta,\phi; \Ish, \DIsh) &:= - \dfrac{\Ish}{2} \cos^2 \theta \sin 2 \phi - \dfrac{ \DIsh}{2} \sin^2 \theta \sin 2 \phi, \\
\ga(\theta,\phi;\Ish) &:= \dfrac{\Ish}{2} \cos \theta \sin 2\phi.
\end{align}
\end{subequations}
Here, $\Ish$ and $\DIsh$ are chirality parameters, where $\Ish$ is sometimes referred to as the Ishimoto parameter \citep{ohmura2021near}. They represent rotational drift due to moments of chirality along the axis of helicoidal symmetry, as summarised in Table \ref{Table: Parameter types}. If the particle is spheroidal, then $\Ish = \DIsh = 0$ and the governing equations for the rotational dynamics reduce to those in Part 1. For brevity, when referring to $f_i$ and $g_i$ we will often suppress the explicit parameter dependence on $\Breth$, $\Ish$, and $\DIsh$ unless specifically relevant. The typical ranges of $\Ish$ and $\DIsh$ are not well explored in the literature for different swimmers, with the notable exception of experimental measurements for bacterial swimmers, giving $\Ish \approx 0.01$ \citep{jing2020chirality, zottl2022asymmetric, ronteix2022rheotaxis}. Given this, for reference we approximate plausible ranges of these parameters for a simple bacterial model using resistive force theory in Appendix \ref{sec: estimations}, which suggest $|\Ish| \approx 0.01$ and $|\DIsh| \approx 0.5$. Since $\psi$ decouples from the system \eqref{eq: full gov eq}--\eqref{eq: g functions} for passive swimmers (i.e. for $\omb = \oma = 0$) and $\Ish$ appears to be small, one might assume that the effects of chirality are unimportant to Jeffery's orbits. We will show that this is not the case in general for the active swimmers we consider. Therefore, we retain both $\Ish$ and $\DIsh$ in our analysis, and we will see that this is important to comprehensively capture the nature of the emergent dynamics.

We now consider the governing equations for $\Xvecpos(t)$, the position of the swimmer in the laboratory frame. While the equivalent equations in Part 1 were fairly intuitive and straightforward to state, this was due to the intrinsic symmetry of spheroidal particles, which removed several of the more general contributions to translation. Since we now consider a more general class of objects, the translational dynamics in Part 2 feature additional contributions. We derive the resulting governing equations of motion in Appendix \ref{sec: derivation eom}, which are
\begin{align}
\label{eq: full gov eq translational}
\dbyd{\Xvecpos}{\tstandard} = \vel + \Ypos \e{3}
- \shapecoeffb \left ( \ehat{2} \ehat{3}^T - \ehat{3} \ehat{2}^T \right ) \mat{E}^* \ehat{1}
+ \shapecoeffc \mat{E}^* \ehat{1}
+ (\shapecoeffa - \shapecoeffc) (\ehat{1}^T\mat{E}^* \ehat{1}) \ehat{1} .
\end{align}
We emphasise that $\vel$ and $\ehat{i}$ depend on the orientation of the object through the Euler angles $(\theta, \psi, \phi)$, which evolve via \eqref{eq: full gov eq}--\eqref{eq: g functions}. The additional terms in \eqref{eq: full gov eq translational} not present in Part 1 involve the rate of strain tensor $\mat{E}^*$, and three additional degrees of freedom encoded through the shape parameters $\shapecoeffb$, $\shapecoeffc$, and $\shapecoeffa$.

These shape parameters can be interpreted as measures of translational drift induced by the coupling between the shear-induced strain and asymmetries in the object shape. As summarised in Table \ref{Table: Parameter types}, $\shapecoeffb$ represents a measure of drift due to chirality of the object, and $\shapecoeffc$, $\shapecoeffa$ represent measures of drift due to fore-aft asymmetry of the object. These additional translational terms arise in a similar manner to the additional terms \eqref{eq: g functions} in the rotational dynamics. If the particle is spheroidal, then $\shapecoeffb = \shapecoeffc = \shapecoeffa = 0$ and the governing equations for the translational dynamics reduce to those in Part 1. In Appendix \ref{sec: estimations}, we estimate typical ranges of the shape parameters $\shapecoeffb$, $\shapecoeffc$, and $\shapecoeffa$ using resistive force theory for a simple model bacterium swimmer.

The full dynamics comprising \eqref{eq: full gov eq}--\eqref{eq: full gov eq translational} govern the motion of any hydrodynamically helicoidal object in shear flow, by definition. That is, as discussed above, helicoidal objects are defined as objects that follow these dynamics, rather than by any necessary geometric properties. However, as we discuss below, there are important sufficient geometric conditions that give rise to hydrodynamic helicoidicity. In the hydrodynamic sense, the behaviour of helicoidal objects in shear flow is therefore fully characterised by the six parameters $\Breth,\Ish,\DIsh,\shapecoeffb,\shapecoeffc,\shapecoeffa$, summarised in Table \ref{Table: Parameter types}. This general class of shapes contains several subclasses of hydrodynamic symmetries, discussed extensively in \citet{Ishimoto2020b}. These subclasses include shapes that possess additional geometric symmetries, and are characterised mathematically by particular combinations of the six shape parameters vanishing.

\begin{table}
\begin{center}
\begin{tabular}{lll}
Parameter      & Type of drift generated                                & Geometric cause        \\ \hline
$\Breth$       & Rotational                                             & Achiral `aspect ratio' \\
$\Ish$         & Rotational component off symmetry axis    & Chiral effects                \\
$\DIsh$        & Rotational component along symmetry axis               & Chiral effects                \\
$\shapecoeffb$ & Translational                                          & Chiral effects                \\
$\shapecoeffc$ & Translational component off symmetry axis & Fore-aft asymmetry     \\
$\shapecoeffa$ & Translational component along symmetry axis            & Fore-aft asymmetry  
\end{tabular}
\caption{Summary of the six parameters that characterise objects with hydrodynamic helicoidal symmetry.}
\label{Table: Parameter types}
\end{center}
\end{table}

\begin{figure}
    \centering
    \includegraphics[width=\textwidth]{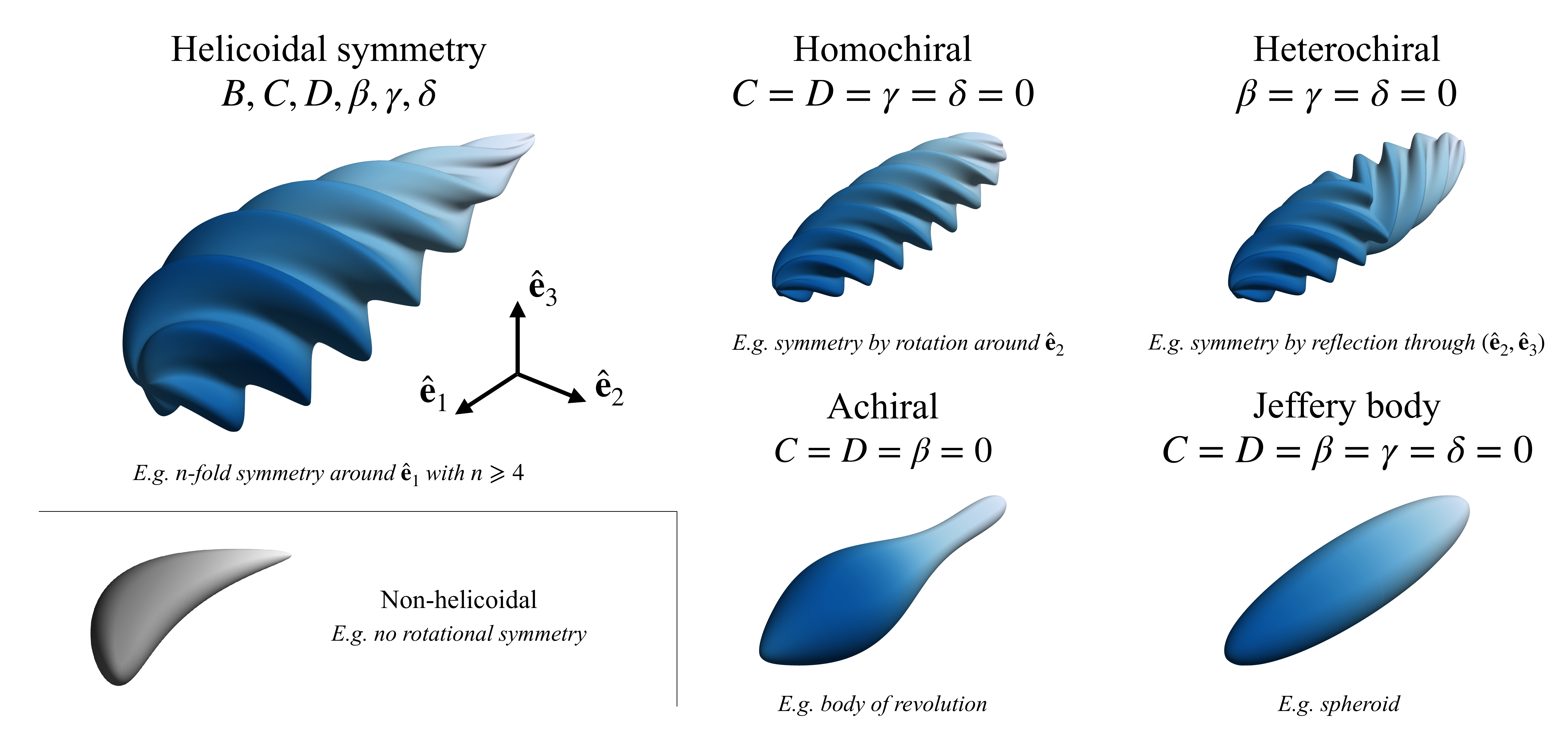}
    \caption{Examples of the class of shapes considered in Part 2. The bodies we investigate possess `helicoidal' symmetry, which allows us to characterise their dynamics in shear flow with six parameters $\Breth,\Ish,\DIsh,\shapecoeffb,\shapecoeffc,\shapecoeffa$. On the right, we distinguish the specific subcases we discuss in the main text: homochirality, heterochirality, achirality and Jeffery bodies, with this last class of objects following the simpler dynamics investigated in Part 1. Each category of shape is illustrated with an example particle possessing additional geometrical symmetries. For comparison, in the bottom left we provide an example of a shape that does not possess helicoidal symmetry and, therefore, is not captured by our analysis.}
    \label{fig:shapes}
\end{figure}

We illustrate an example of a general hydrodynamically helicoidal body in the left panel of Figure \ref{fig:shapes}, recalling that this includes (but is not limited to) objects possessing $n$-fold rotational symmetry along an axis for some integer $n\geqslant 4$. In particular, this allows the object to be chiral and to be free of any fore-aft symmetry constraints.

On the right panel of Figure \ref{fig:shapes}, we illustrate four main hydrodynamic symmetry subcases of interest, giving examples of geometric symmetries that generate the specific subcases. An object that is geometrically symmetric with respect to a rotation of $\pi$ around an axis perpendicular to the helicoidal symmetry axis has $\Ish = \DIsh = \shapecoeffc = \shapecoeffa = 0$; we describe an object satisfying these parameter constraints as possessing \emph{homochiral} hydrodynamic symmetry, following the terminology employed in \cite{Ishimoto2020b}. A homochiral object does not experience any chirality-induced rotational drift. That is, from the governing equations \eqref{eq: full gov eq}--\eqref{eq: full gov eq translational}, the effects of chirality in the dynamics of homochiral objects will only manifest through the drift velocity terms in the translational dynamics \eqref{eq: full gov eq translational}. Thus, the rotational dynamics will remain as classic Jeffery orbits, while the translational dynamics will differ. 

An object that is geometrically symmetric with respect to reflection in a plane normal to the axis of helicoidal symmetry has $\shapecoeffb = \shapecoeffc =\shapecoeffa = 0$; we describe an object satisfying these parameter constraints as possessing \emph{heterochiral} hydrodynamic symmetry. A heterochiral object does not experience any chirality-induced translational drift. In particular, such an object always satisfies $\shapecoeffc = \shapecoeffa = 0$. In contrast to homochiral objects, the effect of chirality in the dynamics of heterochiral objects will appear in the rotational drift terms in the rotational dynamics \eqref{eq: full gov eq}--\eqref{eq: g functions}, resulting in chiral Jeffery orbits. These, in turn, will also influence the translational dynamics \eqref{eq: full gov eq translational}, which are coupled to the evolution of the object orientation. Given the geometric symmetries that generate homochiral or heterochiral objects, we describe an object in either subclass as possessing hydrodynamic \emph{fore-aft symmetry}.

An object that is geometrically symmetric with respect to continuous rotation around the helicoidal axis (i.e. a body of revolution) has $\Ish = \DIsh = \shapecoeffb = 0$; we describe an object satisfying these parameter constraints as possessing \emph{achiral} hydrodynamic symmetry. Similar to the homochiral case, an achiral object does not experience any chirality-induced rotational drift. However, an achiral object will experience a different translational drift to a homochiral object in general.

Finally, any object with at least two of the geometric symmetries described above (e.g. a spheroid) has $\Ish = \DIsh = \shapecoeffb = \shapecoeffc = \shapecoeffa = 0$; as noted in the Introduction, we describe an object satisfying these parameter constraints as a \emph{Jeffery body}. We considered the simpler dynamics of these highly symmetric objects in Part 1.

More generally, in this study we investigate the emergent dynamics of the nonlinear, autonomous dynamical system defined by \eqref{eq: full gov eq}--\eqref{eq: full gov eq translational} for general helicoidal objects in shear flow. In the same manner as Part 1, we consider the regime where the swimmer rotation rate is much larger than the external shear rate. This means that $|\angvel| = |\omb \ehat{1} + \oma \ehat{2}|$ is large. Taking $\omb > 0$ without loss of generality, we consider the distinguished limit where $\oma = \order{\omb}$ with $\omb \gg 1$ (which will give the same information as taking $\abs{\angvel} \gg 1$ with $\angl = \order{1}$), and all other parameters are of $\order{1}$. This asymptotic limit is distinguished in the sense that it is a general case from which the subcases of $|\oma| \gg \omb $ and $\omb \gg |\oma|$ can be distilled as regular asymptotic sublimits of the results we derive.

\section{Setting up the multiple scales analysis}
\label{sec: setup multiscle}

We analyse the system \eqref{eq: full gov eq}--\eqref{eq: full gov eq translational} in the limit of rapid spinning. We animate the full dynamics of this system for various scenarios in Supplementary Movies 1-5. We consider the distinguished limit $\oma = \order{\omb}$ with $\omb \gg 1$ (treating $\omb > 0$ without loss of generality). Given this, it is helpful to introduce the notation $\ac = \order{1}$ such that
\begin{align}
\label{eq: def omega as ratio of Omegas}
\oma = \ac \omb,
\end{align}
and to formally consider the single asymptotic limit $\omb \gg 1$.

Our approach is similar to Part 1; we analyse the system \eqref{eq: full gov eq}--\eqref{eq: full gov eq translational} using the method of multiple scales in the limit of large $\omb$, with the goal of deriving effective equations that govern the emergent behaviour. Moreover, we will see that the leading-order system is equivalent to that of Part 1, so we are able to exploit the multiscale framework we derived therein. The setup for the multiple scales analysis is therefore equivalent to that in Part 1, and we repeat it here for convenience. We reintroduce $\ts$, the fast timescale, via
\begin{align}
\label{eq: new times}
\ts = \left(\oma^2 +\omb^2\right)^{1/2} \tl = \om \omb \tl,
\end{align}
where we use
\begin{align}
\om \coloneqq \sqrt{1 +\ac^2},
\end{align}
for notational convenience, and we refer to the original timescale $\tl$ as the slow timescale. Treating the fast and slow timescales as independent and using \eqref{eq: new times}, the time derivative becomes
\begin{align}
\label{eq: time deriv transform}
\dbyd{}{\tstandard} \mapsto \om \omb \pbyp{}{\ts} + \pbyp{}{\tl}.
\end{align}

Under the time derivative mapping \eqref{eq: time deriv transform}, the rotational dynamics system \eqref{eq: full gov eq} is transformed to
\begin{subequations}
\label{eq: full gov eq trans General}
\begin{align}
\label{eq: theta eq trans General}
\omb \om \pbyp{\theta}{\ts} + \pbyp{\theta}{\tl} &= \omb \ac \cos \psi + \hb(\theta,\phi), \\
\label{eq: psi eq trans General}
\omb \om \pbyp{\psi}{\ts} + \pbyp{\psi}{\tl}  &= \omb\left(1 - \ac \dfrac{\cos \theta \sin \psi}{\sin \theta}\right) + \hc(\theta,\phi), \\
\label{eq: phi eq trans General}
\omb \om  \pbyp{\phi}{\ts} + \pbyp{\phi}{\tl} &= \omb \ac \dfrac{\sin \psi}{\sin \theta} + \ha(\theta,\phi),
\end{align}
\end{subequations}
and the translational dynamics system \eqref{eq: full gov eq translational} is transformed to 
\begin{align}
\om \omb \pbyp{\Xvecpos}{\ts} + \pbyp{\Xvecpos}{\tl} &= \vel + \Ypos \e{3} - \shapecoeffb \left ( \ehat{2} \ehat{3}^T - \ehat{3} \ehat{2}^T \right ) \mat{E}^* \ehat{1} \notag \\
\label{eq: full gov eq translational trans}
&\quad + (\shapecoeffa - \shapecoeffc) (\ehat{1}^T\mat{E}^* \ehat{1}) \ehat{1} + \shapecoeffc \mat{E}^* \ehat{1}.
\end{align}
We expand each dependent variable as an asymptotic series in inverse powers of $\omb$, writing
\begin{align}
\label{eq: asy exp}
\xi(\ts,\tl) \sim \xi_0(\ts,\tl) + \dfrac{1}{\omb} \xi_1(\ts,\tl) \quad \text{as } \omb \to \infty, \quad \text{ for } \xi \in \{\phi, \theta, \psi, X, Y, Z \}.
\end{align}

Since the leading-order (fast) terms in \eqref{eq: full gov eq trans General} and \eqref{eq: full gov eq translational trans} are of $\order{\omb}$, but the new chiral and asymmetric terms are all of $\order{1}$, these new terms do not appear in the leading-order analysis. This means that the leading-order analysis and the adjoint solution used to derive the solvability conditions at next-order are equivalent to those in Part 1, and we can therefore directly use the equivalent results therein. Consequently, we are fairly brief with the leading-order analysis and the derivation of the solvability conditions in the full analysis below, directing the interested reader to Part 1.

\section{Deriving the emergent angular dynamics}
\label{Sec: emergent angular dynamics}

\subsection{Leading-order analysis}
\label{sec: LO}
Using the asymptotic expansions \eqref{eq: asy exp} in the transformed governing equations \eqref{eq: full gov eq trans General}, we obtain the leading-order (i.e. $\order{\omb}$) system
\begin{subequations}
\label{eq: full gov eq trans LO General}
\begin{align}
\label{eq: theta eq trans LO General}
\om \pbyp{\theta_0}{\ts}  &= \ac\cos \psi_0, \\
\label{eq: psi eq trans LO General}
\om \pbyp{\psi_0}{\ts}  &= 1 - \ac \dfrac{\cos \theta_0 \sin \psi_0}{\sin \theta_0}, \\
\label{eq: phi eq trans LO General}
\om \pbyp{\phi_0}{\ts} &= \ac\dfrac{\sin \psi_0}{\sin \theta_0}.
\end{align}
\end{subequations}
We show in \S4.1 of Part 1 that the solution to the nonlinear system \eqref{eq: full gov eq trans LO General} is:
\begin{subequations}
\label{eq: mu def}
\begin{align}
\label{eq: om cos thet def}
\om \cos \theta_0 &= \cos \alp - \ac \sin \alp \cos (\ts + \muc), \\
\label{eq: sin theta sin psi}
\om \sin \theta_0 \sin \psi_0 &= \ac \cos \alp + \sin \alp \cos (\ts + \muc), \\
\label{eq: tan phi0}
\tan \phi_0 &= \dfrac{\ac \cos \phic \sin (\ts + \muc) + \sin \phic \left(\ac \cos \alp \cos (\ts + \muc) + \sin \alp \right)}{\cos \phic \left(\ac \cos \alp \cos (\ts + \muc) + \sin \alp \right) - \ac \sin \phic \sin (\ts + \muc)},
\end{align}
 where $\alp = \alp(\tl)$, $\muc = \muc(\tl)$, and $\phic = \phic(\tl)$ are the three slow-time functions of integration that remain undetermined from our leading-order analysis. The goal of the next-order analysis in \S \ref{sec: FC} is to derive the governing equations satisfied by $\alp$, $\muc$, and $\phic$. As in Part 1, one can think of $\alp$ as controlling some emergent amplitude of oscillation, $\muc$ as controlling some emergent phase of oscillation, and $\phic$ as the emergent drift in yawing. We will also show later that $\alp$ can be associated with $\theta$, $\muc$ with $\psi$, and $\phic$ with $\phi$.

Before proceeding, it will be helpful later to note the additional relationships
\begin{align}
\label{eq: sin theta cos psi}
\sin \theta_0 \cos \psi_0 &= -\sin \alp \sin (\ts + \muc), \\
\label{eq: sin sq theta}
\om^2 \sin^2 \theta_0 &= (\ac \cos \alp \cos (\ts + \muc) + \sin \alp)^2 +  \ac^2\sin^2 (\ts + \muc),
\end{align}
\end{subequations}
where the former follows from differentiating \eqref{eq: om cos thet def} with respect to $\ts$ and imposing \eqref{eq: theta eq trans LO General}, and the latter follows from rearranging \eqref{eq: om cos thet def}.

\subsection{Next-order system}
\label{sec: FC}

Our remaining goal is to determine the governing equations satisfied by the slow-time functions $\alp(\tl)$, $\muc(\tl)$, and $\phic(\tl)$. To do this, we must determine the solvability conditions required for the first-order correction (i.e. $\order{1}$) terms in \eqref{eq: full gov eq trans General} after posing the asymptotic expansions \eqref{eq: asy exp}. These $\order{1}$ terms are
\begin{subequations}
\label{eq: full gov eq trans oma O1 eps}
\begin{align}
\label{eq: theta eq trans oma O1 eps}
\om \pbyp{\theta_1}{\ts} + \ac \psi_1 \sin \psi_0  &= \hb(\theta_0,\phi_0) - \pbyp{\theta_0}{\tl}, \\
\label{eq: psi eq trans oma O1 eps}
\om \pbyp{\psi_1}{\ts} - \ac \theta_1 \dfrac{\sin \psi_0}{\sin^2 \theta_0} + \ac \psi_1 \dfrac{\cos \theta_0 \cos \psi_0}{\sin \theta_0} &= \hc(\theta_0,\phi_0) - \pbyp{\psi_0}{\tl}, \\
\label{eq: phi eq trans oma O1 eps}
\om \pbyp{\phi_1}{\ts} + \ac \theta_1 \dfrac{\cos \theta_0 \sin \psi_0}{\sin^2 \theta_0} - \ac \psi_1\dfrac{\cos \psi_0}{\sin \theta_0} &=  \ha(\theta_0,\phi_0) - \pbyp{\phi_0}{\tl},
\end{align}
\end{subequations}
along with $2 \pi$-periodicity in $\ts$. The system \eqref{eq: full gov eq trans oma O1 eps} constitutes a non-autonomous linear coupled 3D system for $(\theta_1, \psi_1, \phi_1)$ with an inhomogeneous forcing in terms of the leading-order solution.

To derive the required solvability conditions, we use the method of multiple scales for systems (see, for example, pp.~127--128 of \citet{dalwadi2014flow} or p.~22 of \citet{dalwadi2018effect}). As detailed in \S4.2 of Part 1, this entails taking the dot product of the vector solution to the homogeneous adjoint version of \eqref{eq: full gov eq trans oma O1 eps} with the vector right-hand side of \eqref{eq: full gov eq trans oma O1 eps}, and averaging over one fast-time oscillation. We calculate the adjoint solution in \S4.3 and Appendix D of Part 1; using this to apply the procedure outlined above yields the following three solvability conditions
\begin{subequations}
\label{eq: solv conditions transformed}
\begin{align}
- \om \sin \alp \dbyd{\alp}{\tl} &= \dfrac{\om \Beff}{2} \sin^2 \alp \cos \alp \sin 2 \phic \notag \\
&\quad + \av{\gb \left(\ac \cos \theta_0 \sin \psi_0 - \sin \theta_0 \right) + \gc \ac \sin \theta_0 \cos \psi_0}, \\
\om \sin^2 \alp \dbyd{\muc}{\tl} &= \dfrac{\om \Beff}{2} \sin^2 \alp \cos \alp \cos 2 \phic\notag \\
&\quad + \av{\gb \ac \cos \psi_0 + \gc \sin \theta_0 \left( \sin \theta_0 - \ac \cos \theta_0 \sin \psi_0 \right)}, \\
\cos \alp \dbyd{\muc}{\tl} + \dbyd{\phic}{\tl} &= \dfrac{1}{2}\left(1 - \Beff \sin^2 \alp \cos 2 \phic \right) + \av{\gc \cos \theta_0 + \ga},
\end{align}
\end{subequations}
where we have used the results from \S 4.2--4.3 of Part 1 to evaluate all the non-chiral terms (i.e. all terms not involving $g_i$), including the use of the effective Bretherton parameter we derived in Part 1:
\begin{align}
\label{eq: effective B}
\Beff := \dfrac{(2 - \ac^2)\Breth}{2 \left(1 + \ac^2 \right)}.
\end{align}
Additionally, we
use the notation $\av{\bcdot}$ to denote the average of its argument over one fast-time oscillation, explicitly defining
\begin{align}
\label{eq: av operator}
\av{y} = \dfrac{1}{2 \pi}\int_0^{2\pi} \! y \, \mathrm{d}\ts.
\end{align}

Our remaining task is to evaluate the outstanding averages in the three solvability conditions \eqref{eq: solv conditions transformed}, each of which involves the chiral contributions $g_i$ defined in \eqref{eq: g functions}. We have explicit representations of the terms involving the trigonometric functions of $\theta_0$ and $\psi_0$ through the leading-order solutions \eqref{eq: mu def}. The terms involving $\sin 2 \phi_0$ and $\cos 2 \phi_0$, which arise from the $g_i$ defined in \eqref{eq: f functions}, require additional calculation. To derive expressions for these double-angled quantities, we first note
\begin{subequations}
\label{eq: a 0 LO identities General a}
\begin{align}
\label{eq: a 0 LO identities General a sin thet sin phi}
\om \sin \theta_0 \sin \phi_0 &= \ac \cos \phic \sin \sig + \sin \phic \left(\ac \cos \alp \cos \sig + \sin \alp \right), \\
\om \sin \theta_0 \cos \phi_0 &= \cos \phic \left(\ac \cos \alp \cos \sig + \sin \alp \right) - \ac \sin \phic \sin \sig,
\end{align}
\end{subequations}
using the shorthand $\sig = \ts + \muc$. The expressions \eqref{eq: a 0 LO identities General a} are calculated via the identities $\tan \phi_0 = (\om \sin \theta_0 \sin \phi_0)/(\om \sin \theta_0 \cos \phi_0)$, and $\om^2 \sin^2 \theta_0 = (\om \sin \theta_0 \sin \phi_0)^2 + (\om \sin \theta_0 \cos \phi_0)^2$, the left-hand sides of which are defined in \eqref{eq: tan phi0}, \eqref{eq: sin sq theta}. Then, from the expressions of \eqref{eq: a 0 LO identities General a}, appropriate double-angle formulae imply that
\begin{subequations}
\label{eq: double angle relationships}
\begin{align}
\label{eq: cos 2 phi}
\om^2 \sin^2 \theta_0 \cos 2 \phi_0 &= \Ceven(\sig,\tl) \cos 2\phic - \Sodd(\sig,\tl) \sin 2 \phic, \\
\label{eq: sin 2 phi}
\om^2 \sin^2 \theta_0  \sin 2 \phi_0 &= \Sodd(\sig,\tl) \cos 2\phic + \Ceven(\sig,\tl) \sin 2 \phic, \\
\label{eq: Ceven and Sodd}
\Ceven(\sig,\tl) &:= \left(\ac \cos \alp \cos \sig + \sin \alp\right)^2 - \ac^2 \sin^2\sig, \\
\label{eq: Sodd}
\Sodd(\sig,\tl) &:= 2 \ac \sin\sig \left(\ac \cos \alp \cos \sig + \sin \alp \right).
\end{align}
\end{subequations}

We can now simplify the remaining fast-time averages in the right-hand side of \eqref{eq: solv conditions transformed}. We start by exploiting the parity of various expressions. Specifically, we use the evenness of $\cos \theta_0$, $\sin \theta_0 \sin \psi_0$, $\sin^2 \theta_0$, and $\Ceven$ around $\sig = \pi$ (from \eqref{eq: om cos thet def}, \eqref{eq: sin theta sin psi}, \eqref{eq: sin sq theta}, and \eqref{eq: Ceven and Sodd}, respectively), and the oddness of $\sin \theta_0 \cos \psi_0$ and $\Sodd$ around $\sig = \pi$ (from \eqref{eq: sin theta cos psi} and \eqref{eq: Sodd}, respectively). This allows us to write the fast-time averages in the right-hand side of \eqref{eq: solv conditions transformed} as
\begin{subequations}
\label{eq: g_i RHS simplify}
\begin{align}
&\av{\gb \left(\ac \cos \theta_0 \sin \psi_0 - \sin \theta_0 \right) + \gc \ac \sin \theta_0 \cos \psi_0}\notag \\
&\qquad = -\dfrac{\Ish}{2 \om^2} \cos 2 \phic \av{\Ceven \left( \dfrac{\ac \cos \theta_0 \sin \psi_0}{\sin \theta_0} - 1\right)+ \dfrac{\Sodd \ac \cos \psi_0}{\sin \theta_0}} \notag \\
&\qquad \qquad +\dfrac{\Ish - \DIsh}{2 \om^2} \cos 2 \phic \av{\Sodd \ac \sin \theta_0 \cos \psi_0}, \\
&\av{\gb\ac \cos \psi_0 + \gc \sin \theta_0 \left( \sin \theta_0 - \ac \cos \theta_0 \sin \psi_0 \right)} \notag \\
&\qquad = \dfrac{\Ish}{2 \om^2} \sin 2 \phic \av{\Ceven \left(\dfrac{\ac \cos \theta_0 \sin \psi_0}{\sin \theta_0} - 1 \right) +  \dfrac{\Sodd \ac \cos \psi_0}{\sin \theta_0}} \notag \\
&\qquad \qquad + \dfrac{\DIsh - \Ish}{2 \om^2} \sin 2 \phic \av{\Ceven \left( \ac \cos \theta_0 \sin \theta_0 \sin \psi_0 - \sin^2 \theta_0 \right)}, \\
&\av{\gc \cos \theta_0 + \ga} 
= \dfrac{\Ish - \DIsh}{2 \om^2} \sin 2 \phic \av{\Ceven \cos \theta_0}.
\end{align}
\end{subequations}
Using the leading-order solutions \eqref{eq: mu def} with the definitions of $\Ceven$ and $\Sodd$ in \eqref{eq: Ceven and Sodd}--\eqref{eq: Sodd}, we may write the terms within the averages on the right-hand sides of \eqref{eq: g_i RHS simplify} explicitly as
\begin{subequations}
\label{eq: g_i RHS simplify in sig}
\begin{align}
&\Ceven \left( \dfrac{\ac \cos \theta_0 \sin \psi_0}{\sin \theta_0} - 1\right)+ \dfrac{\Sodd \ac \cos \psi_0}{\sin \theta_0} = -\left(1 + \ac^2 \right)\sin \alp \left( \sin \alp + \ac \cos \alp \cos \sig \right), \\
&\Sodd \ac \sin \theta_0 \cos \psi_0 = -2 \ac^2 \sin \alp \sin^2 \sig \left( \sin \alp + \ac \cos \alp \cos \sig \right) , \\
&\Ceven \left(\ac \cos \theta_0 \sin \theta_0 \sin \psi_0  - \sin^2 \theta_0 \right) \notag\\
&\quad= \sin \alp \left( \sin \alp + \ac \cos \alp \cos \sig \right) \left(\ac^2 \sin^2 \sig - \left( \sin \alp + \ac \cos \alp \cos \sig \right)^2 \right), \\
&\Ceven \cos \theta_0 = \dfrac{\cos \alp - \ac \sin \alp \cos \sig}{\om}\left(\left(\sin \alp + \ac \cos \alp \cos \sig \right)^2 - \ac^2 \sin^2\sig \right).
\end{align}
\end{subequations}
We can now explicitly calculate the averages of the right-hand sides of \eqref{eq: g_i RHS simplify in sig} over one fast-time oscillation, to deduce that
\begin{subequations}
\label{eq: g_i RHS simplify averages}
\begin{align}
\av{\Ceven \left( \dfrac{\ac \cos \theta_0 \sin \psi_0}{\sin \theta_0} - 1\right)+ \dfrac{\Sodd \ac \cos \psi_0}{\sin \theta_0}} &= -(1 + \ac^2)\sin^2 \alp, \\
\av{\Sodd \ac \sin \theta_0 \cos \psi_0} &= -\ac^2 \sin^2 \alp, \\
\av{\Ceven \left(\ac \cos \theta_0 \sin \theta_0 \sin \psi_0  - \sin^2 \theta_0 \right)} &= -\dfrac{\sin^2 \alp}{2} \left(2 \ac^2 \cos^2 \alp + (2 - \ac^2) \sin^2 \alp \right), \\
\av{\Ceven \cos \theta_0} &= \dfrac{2 - 3 \ac^2}{2 \om} \cos \alp \sin^2 \alp.
\end{align}
\end{subequations}
Substituting \eqref{eq: g_i RHS simplify averages} into \eqref{eq: g_i RHS simplify}, we deduce the following expressions for the averages of the chiral terms
\begin{subequations}
\label{eq: g_i RHS}
\begin{align}
&\av{\gb \left(\ac \cos \theta_0 \sin \psi_0 - \sin \theta_0 \right) + \gc \ac \sin \theta_0 \cos \psi_0} = \dfrac{\Ish + \ac^2 \DIsh}{2 \om^2} \cos 2 \phic \sin^2 \alp, \\
&\av{\gb\ac \cos \psi_0 + \gc \sin \theta_0 \left( \sin \theta_0 - \ac \cos \theta_0 \sin \psi_0 \right)} \notag \\
&\, = -\dfrac{1}{2 \om^2} \sin 2 \phic \sin^2 \alp \left( \left(\Ish + \ac^2 \DIsh \right) \cos^2 \alp + \dfrac{3 \ac^2 \Ish + (2 - \ac^2)\DIsh}{2} \sin^2 \alp \right), \\
&\av{\gc \cos \theta_0 + \ga} 
= \dfrac{(\Ish - \DIsh) \left(2  - 3\ac^2\right)}{4 \om^2} \sin 2 \phic \cos \alp \sin^2 \alp.
\end{align}
\end{subequations}

\begin{figure}
\begin{center}
\vspace{0.5em}
\begin{overpic}[permil,width=\textwidth]{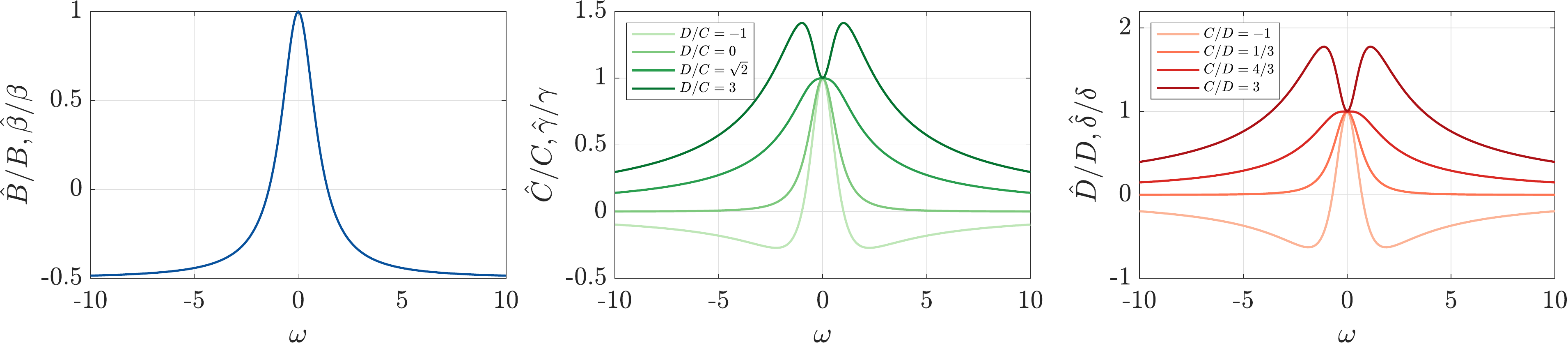}
    \put(10,215){(a)}
    \put(332,215){(b)}
    \put(682,215){(c)}
\end{overpic}
\caption{The effective parameters $\Beff, \Ceff, \Deff, \effshapecoeffb, \effshapecoeffc, \effshapecoeffa$ as functions of $\ac$, normalised by their intrinsic equivalents. (a) $\Beff$ and $\effshapecoeffb$ are only functions of $\ac$, and exhibit the same dependence on $\ac$ following normalisation. (b,c) The remaining effective parameters are functions of three parameters. All are coupled to $\ac$, the orientational shape parameters are also coupled to $\Ish$ and $\DIsh$, while the translational shape parameters are also coupled to $\shapecoeffc$ and $\shapecoeffa$ instead. We show selected curves for different parameter values. Several of the effective coefficients display non-trivial zeros as functions of $\ac$. This suggests that specific activity-induced spinning can effectively eliminate certain parameters, and hence the associated physical interactions of an object with the flow.}
\label{fig: bhat chat dhat}
\end{center}
\end{figure}

Finally, to obtain the slow-time governing equations for $\alp$, $\muc$, and $\phic$ that we have been seeking, we substitute the explicit averages \eqref{eq: g_i RHS} into the solvability conditions \eqref{eq: solv conditions transformed}, and rearrange to obtain the following reduced system
\begin{subequations}
\label{eq: original reduced}
\begin{align}
\dbyd{\alp}{\tl} &= -\dfrac{\Beff}{2} \sin \alp \cos \alp \sin 2 \phic - \dfrac{\Ceff}{2} \sin \alp \cos 2 \phic, \\
\dbyd{\muc}{\tl} &= \dfrac{\Beff}{2} \cos \alp \cos 2 \phic - 
\dfrac{\Ceff}{2} \cos^2 \alp \sin 2 \phic - \dfrac{\Deff}{2} \sin^2 \alp \sin 2 \phic, \\
\dbyd{\phic}{\tl} &= \dfrac{1}{2}\left(1 - \Beff \cos 2 \phic \right) + \dfrac{\Ceff}{2} \cos \alp \sin 2 \phic,
\end{align}
\end{subequations}
where we define the effective chiral coefficients 
\begin{align}
\label{eq: effective coefficients}
\Ceff := \dfrac{\Ish + \ac^2 \DIsh}{(1 + \ac^2)^{3/2}} , \qquad
\Deff := \dfrac{3 \ac^2 \Ish + (2 - \ac^2) \DIsh}{2 (1 + \ac^2)^{3/2}}.
\end{align}
We illustrate these effective parameters in terms of $\ac$ in Figure \ref{fig: bhat chat dhat}.

\subsubsection{Summary}
\label{sec: Summary orientational dynamics}
By comparison with the original angular dynamical system, defined in \eqref{eq: full gov eq}--\eqref{eq: g functions}, we see that the emergent dynamics governed by \eqref{eq: original reduced} can be re-written in terms of the combined achiral and chiral functions $h_i = f_i + g_i$ as follows
\begin{align}
\label{eq: simplified reduced}
\dbyd{\alp}{\tl} = \hb(\alp,\phic; \Beff, \Ceff), \quad
\dbyd{\muc}{\tl} = \hc(\alp,\phic;\Beff, \Ceff, \Deff), \quad
\dbyd{\phic}{\tl} = \ha(\alp,\phic;\Beff, \Ceff),
\end{align}
where the effective Bretherton parameter $\Beff$ is defined in \eqref{eq: effective B}, and the effective chiral coefficients $\Ceff$ and $\Deff$ are defined in \eqref{eq: effective coefficients}.

Therefore, similar to Part 1, the emergent dynamics for rapidly spinning chiral particles are governed by a system that has the same functional form as the original dynamical system without rapid spinning, but with modified coefficients \eqref{eq: effective coefficients} that account for the effect of the spinning. As before, we can identify each slow-time function with an underlying variable: $\alp$ with $\theta$, $\muc$ with $\psi$, and $\phi$ with $\phic$. Since the slow terms in the original dynamical system represent the generalised Jeffery's equations for chiral particles, we can say that rapidly spinning chiral particles behave as particles with an effective chirality, as quantified through the effective coefficients \eqref{eq: effective coefficients}.

\begin{figure}
\begin{center}
\begin{overpic}[permil,width=\textwidth]{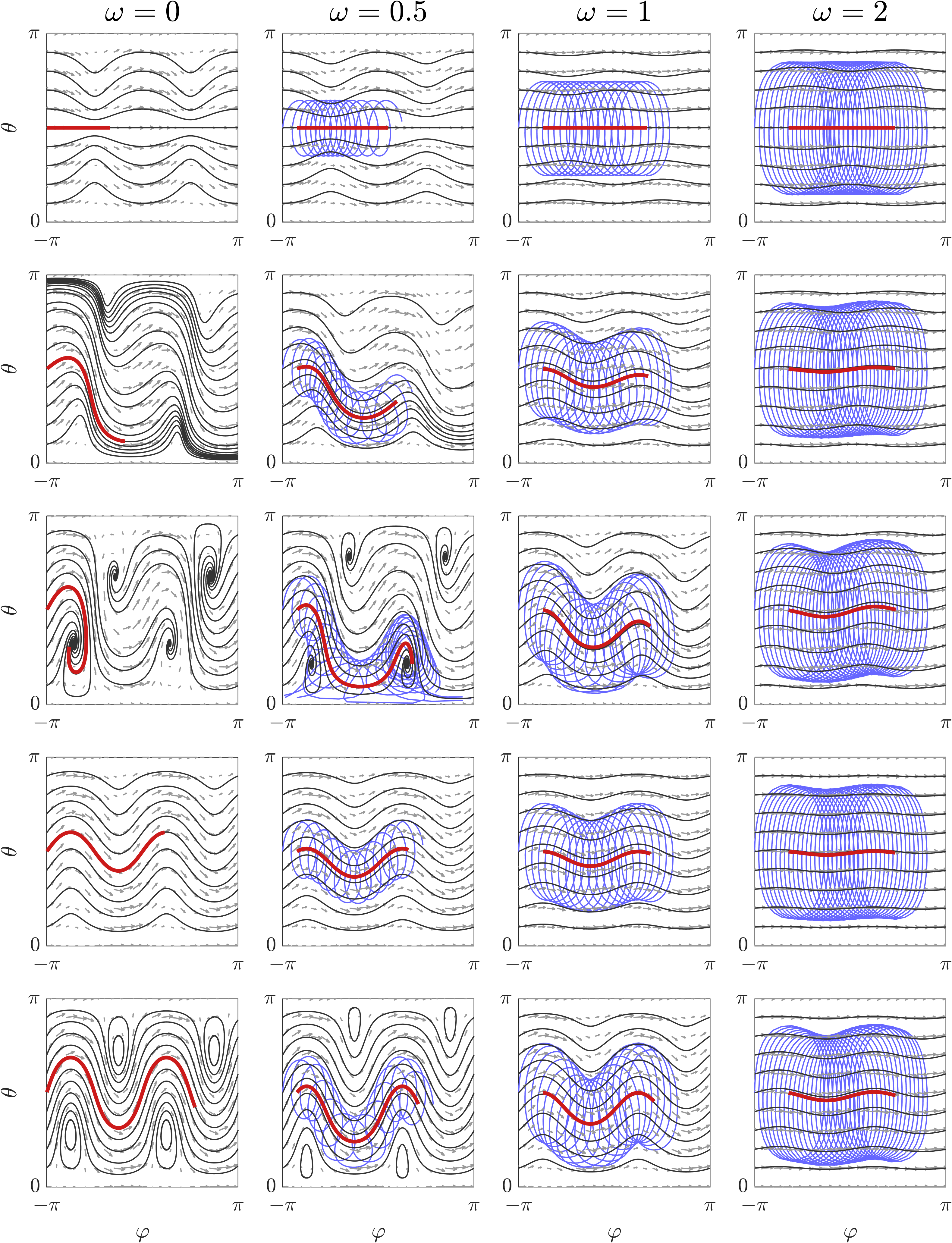}
    \put(-10,980){(a)}
    \put(-10,785){(b)}
    \put(-10,590){(c)}
    \put(-10,395){(d)}
    \put(-10,200){(e)}
\end{overpic}
\caption{Exploring the orientational dynamics in the $(\phi,\theta)$-plane for various values of $\Breth$, $\Ish$ and $\ac$, with sample, rapidly oscillating full dynamics shown in blue for $\ac\neq0$, and the corresponding averaged dynamics shown in red. (a) $(\Breth,\Ish) = (0.7,0)$. (b) $(\Breth,\Ish) = (0.7,0.7)$. (c) $(\Breth,\Ish) = (0.7,1.5)$. (d) $(\Breth,\Ish) = (0,0.7)$. (e) $(\Breth,\Ish) = (0,1.5)$. We use $\DIsh=0$ and $(\theta,\phi) = (\pi/2,-\pi)$ at initial time throughout. For the blue lines, we also set $\omb = 10$ and $\oma = 10 \ac$. Dynamic versions of the full dynamics of the highlighted trajectories in rows (b)-(e) are given in Supplementary Movies 1-4.}
\label{fig:flow}
\end{center}
\end{figure}

We explore the effect of rotation on the orientational dynamics in Figure \ref{fig:flow} and Supplementary Movies 1-4. In Figure \ref{fig:flow}, we illustrate trajectories in the $(\phi,\theta)$-plane and set $\DIsh = 0$ for simplicity. In Figure \ref{fig:flow}a-c, we fix a Bretherton parameter of $\Breth = 0.7$ and vary the chirality parameter $\Ish$ in order to highlight the qualitative changes that chirality can induce. In the first row (a), we set $\Ish = 0$ and present standard Jeffery orbits for homochiral particles for the purpose of comparison, which are periodic as $\abs{\Breth} < 1$. Since this sublimit is a regular limit of the achiral analysis of Part 1, the trajectories shown in this row are identical to those explored in Part 1. In the second row (b), we increase the chirality parameter to $\Ish = 0.7$, illustrating the trajectories of chiral objects. Here, the chirality breaks the periodicity of the slow-time generalised Jeffery trajectories for smaller values of $\ac$, instead inducing a drift towards the pole $\theta=0$. However, this periodicity-breaking effect appears to weaken for larger values of $\ac$, when the effective chirality $\Ceff$ of the object is reduced following \eqref{eq: effective coefficients}. In the third row (c), we show trajectories for a strongly chiral object, increasing the chirality parameter to $\Ish = 1.5$. Here, the chirality induced periodicity-breaking effect is stronger, with the notable appearance of attractive and repulsive points away from the poles at $\theta = 0,\pi$, and persists for larger values of $\ac$ before eventually leading to approximately periodic trajectories as $\ac$ further increases.

\begin{figure}
\begin{center}
\begin{overpic}[permil,width=\textwidth]{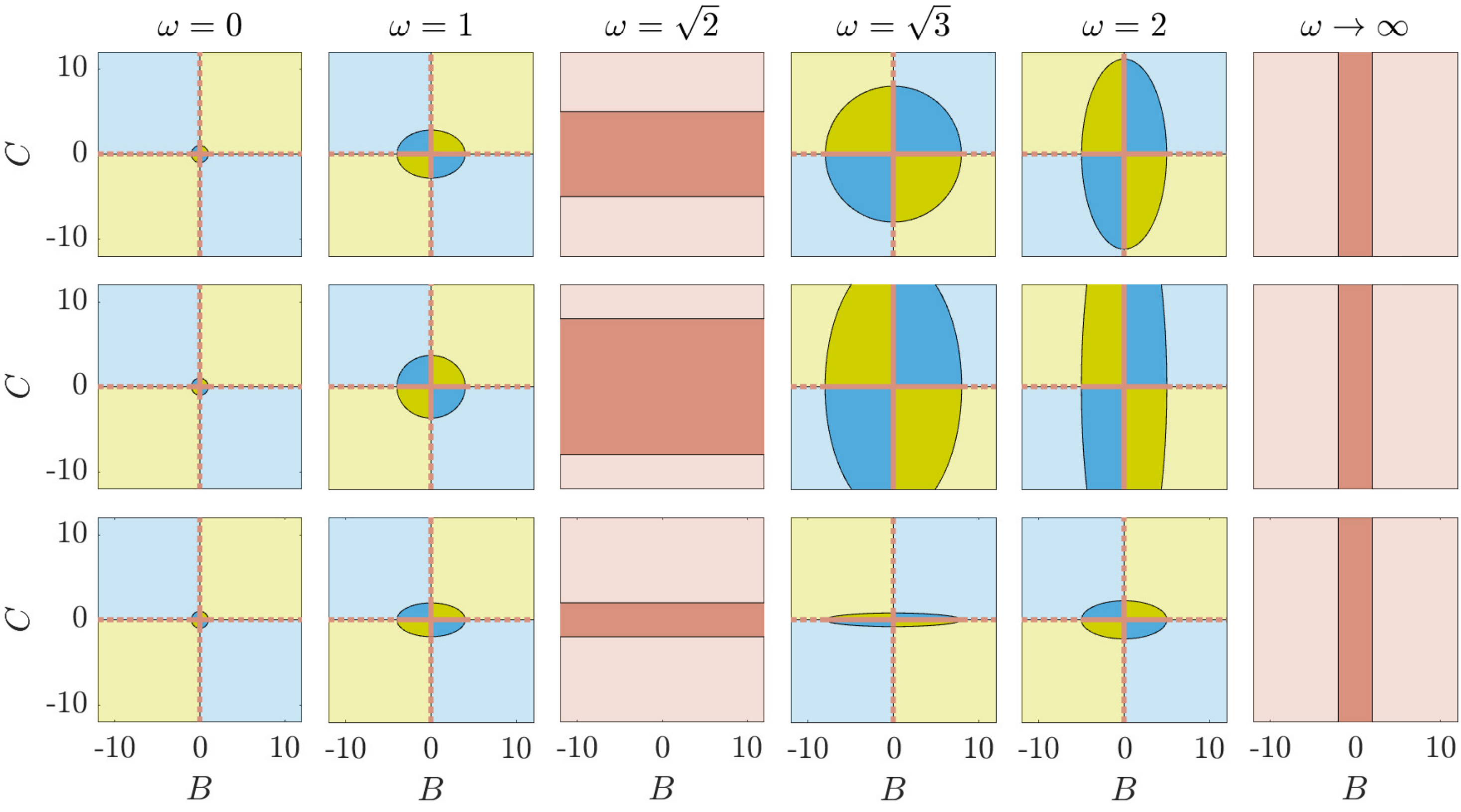}
    \put(-5,495){(a)}
    \put(-5,335){(b)}
    \put(-5,180){(c)}
\end{overpic}
\caption{Schematic showing the qualitative nature of the orientational dynamics within the parameter space $(\Breth,\Ish)$, for different values of $\ac$. The darker regions within each ellipse indicate that trajectories drift towards a pole ($\theta = 0$ for yellow region, $\theta = \pi$ for blue region). Outside the ellipses, the pole solutions become repulsive points and non-trivial attractors exist. The lighter regions external to each ellipse indicate that these non-trivial attractors are in the northern (yellow) and southern (blue) hemispheres, respectively. The thicker red lines (solid and dashed) on the axes indicate periodic trajectories. Dashed lines indicate the existence of orbits that are not centered around one of the poles at $\theta = 0, \pi$. In the critical cases $\ac= \sqrt{2}$ and $\ac \to \infty$, all trajectories are orbits, so there only exist red regions. There is a distinction between orbits centred around a pole (darker red) and not centred around a pole (lighter red). Finally, the influence of the third shape parameter $\DIsh$ is shown in each row: (a) $\DIsh=0$, (b) $\DIsh=0.3$, (c) $\DIsh=-0.3$.
}
\label{fig: bifuraction diagram}
\end{center}
\end{figure}

In Figure \ref{fig:flow}d,e, we consider the effects of chirality on an object with vanishing Bretherton parameter, setting $\Breth = 0$. In (d), we take $\Ish = 0.7$, observing periodic trajectories whose behaviour is significantly more oscillatory in the $\theta$ variable than in the classical Jeffery orbits of Figure \ref{fig:flow}a. Further, $\theta = \pi/2$ is no longer a steady solution, which can also be seen by directly considering the contribution of the chiral function $g_1$ of \eqref{eq: g functions 1} in the governing equation \eqref{eq: theta eq}. As $\ac$ increases, we see a general reduction in these oscillations towards those of a sphere (with $\Breth=\Ish=0$), as predicted by our explicit result for the effective chirality $\Ceff$ in \eqref{eq: effective coefficients}. In (e), the fifth and final row, we consider a strongly chiral object by taking $\Ish = 1.5$. In this case, the strongly chiral effects induce periodic orbits that, curiously, do not encircle the pole for smaller values of $\ac$, instead orbiting around non-trivial fixed points in the $(\phi,\theta)$-plane. However, as $\ac$ increases and decreases $\Ceff$, these orbits collapse, and the trajectories begin to approach those seen in Figure \ref{fig:flow}d for smaller values of $\ac$, as expected. The existence of periodic orbits that do not encircle the pole for larger values of $\Ish$ is due to the pole becoming a repulsive fixed point when $\Breth^2+\Ish^2 > 1$ in the case of a passive object \citep{Ishimoto2020a, Ishimoto2020b}, with non-trivial attractors emerging as a result of the bifurcation. In Figure \ref{fig: bifuraction diagram}, we provide a visual characterisation of the qualitative behaviour of the solution space for the orientational dynamics in terms of the effective parameters $\Beff$ and $\Ceff$.

Given these observations, it is of interest to note the limiting cases of $\ac \to 0$ and $|\ac| \to \infty$. In the limit of $\ac \to 0$, the effective chiral parameters remain the same i.e. $\Ceff \to \Ish$ and $\Deff \to \DIsh$. That is, when spinning is rapid only around the axis of helicoidal symmetry, the effective shape of the chiral swimmer is unchanged; the rapid rotation does not significantly impact the emergent angular dynamics. On the other hand, in the limit of $|\ac| \to \infty$, the effective chiral parameters vanish i.e. $\Ceff \to 0$ and $\Deff \to 0$. That is, when rapid spinning only around an axis perpendicular to the axis of helicoidal symmetry, the rapid rotation causes the chiral swimmer to lose the effect of its chirality and for its orientation to evolve as though it were an achiral particle. This is because the coefficients $\Ish$ and $\DIsh$ can be thought of as moments of chirality along the axis of helicoidal symmetry, and rapid rotation around an axis perpendicular to this will `spread out' the chirality on average, reducing the effective moment to zero. 

Additionally, we see that a chiral particle with $\Ish = 0$ but $\DIsh \neq 0$  (or $\DIsh = 0$ but $\Ish \neq 0$) can result in $\Ceff \neq 0$ and $\Deff \neq 0$. That is, in certain cases with chiral particles, rapid spinning can generate effective terms that were not present in the original equations. Moreover, rapid spinning can either enhance or diminish the effects of chirality, depending on the specific values of $\Ish$ and $\DIsh$ and the relative rotation ratio $\ac$.

\begin{figure}
    \centering
    \vspace{3em}
    \hspace{-2em}\begin{overpic}[width=4.9cm]{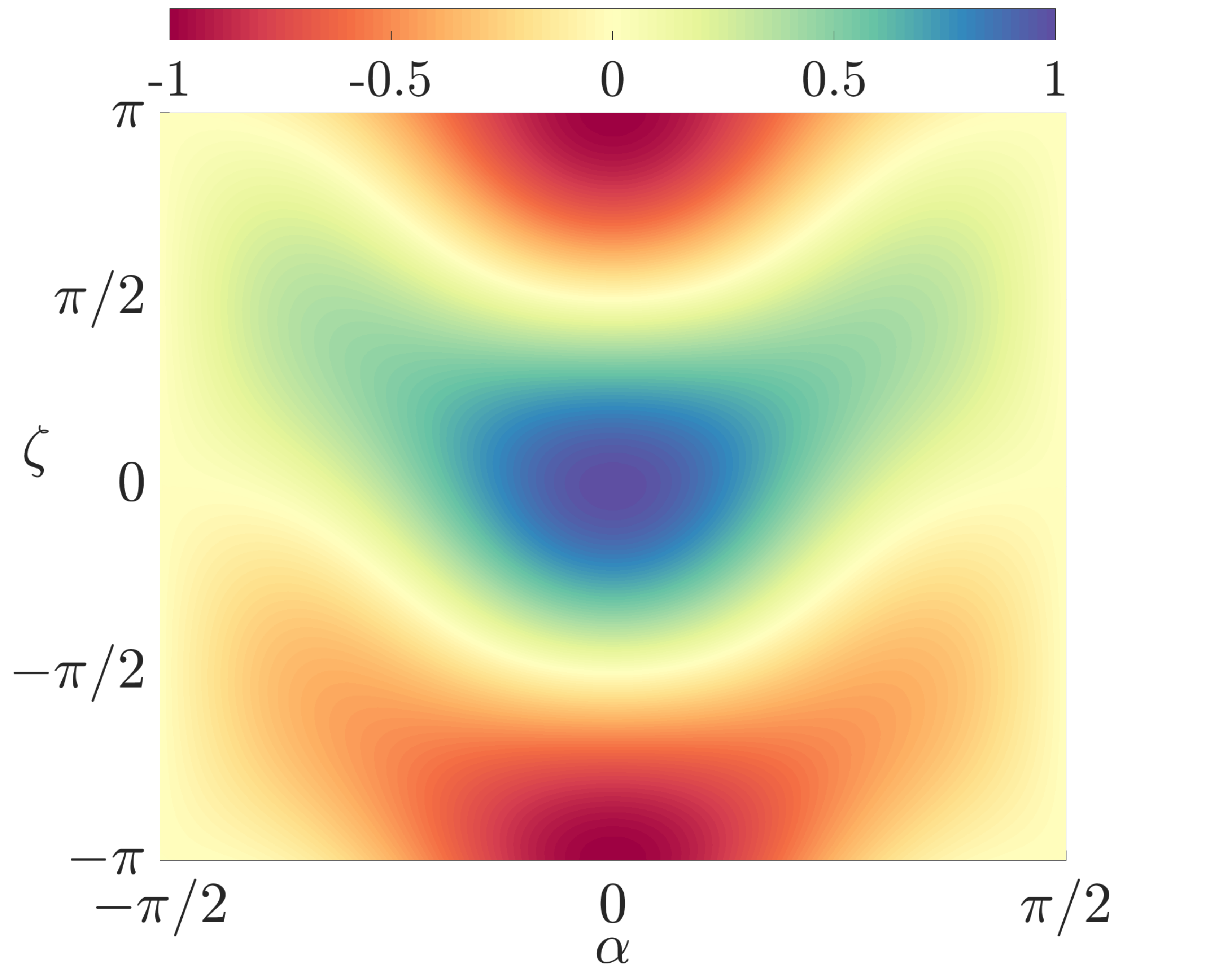}
    \put(0,88){(a)}
    \put(35,85){ \scalebox{.7}{$\dfrac{\Ceff}{\sqrt{\Ish^2 + \DIsh^2}}$} }
    \end{overpic}
         \hspace{-2em}
    \begin{overpic}[width=4.9cm]{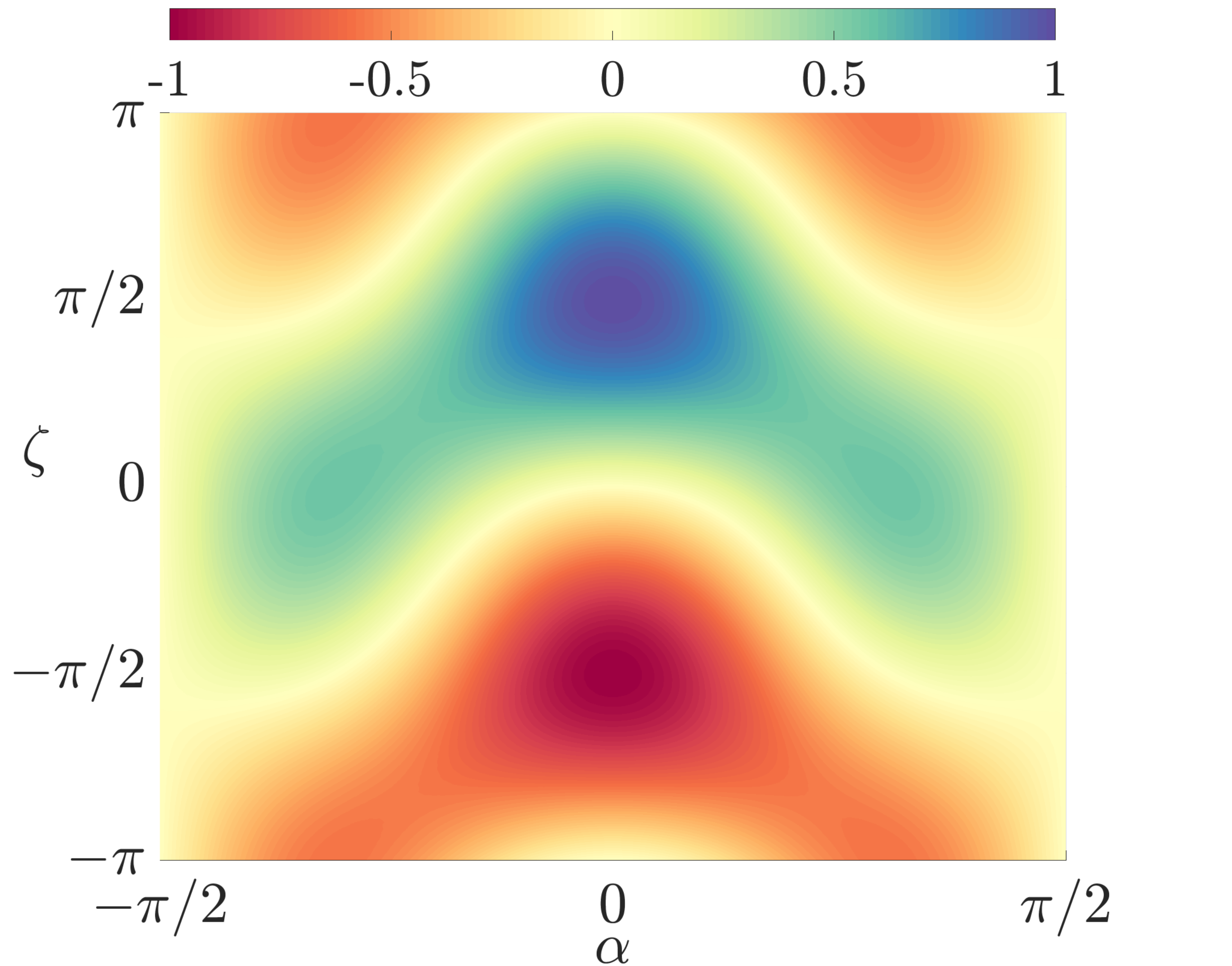}
    \put(0,88){(b)}
    \put(35,85){ \scalebox{.7}{$\dfrac{\Deff}{\sqrt{\Ish^2 + \DIsh^2}}$} }
    \end{overpic}  
    \hspace{-2em}
        \begin{overpic}[width=4.9cm]{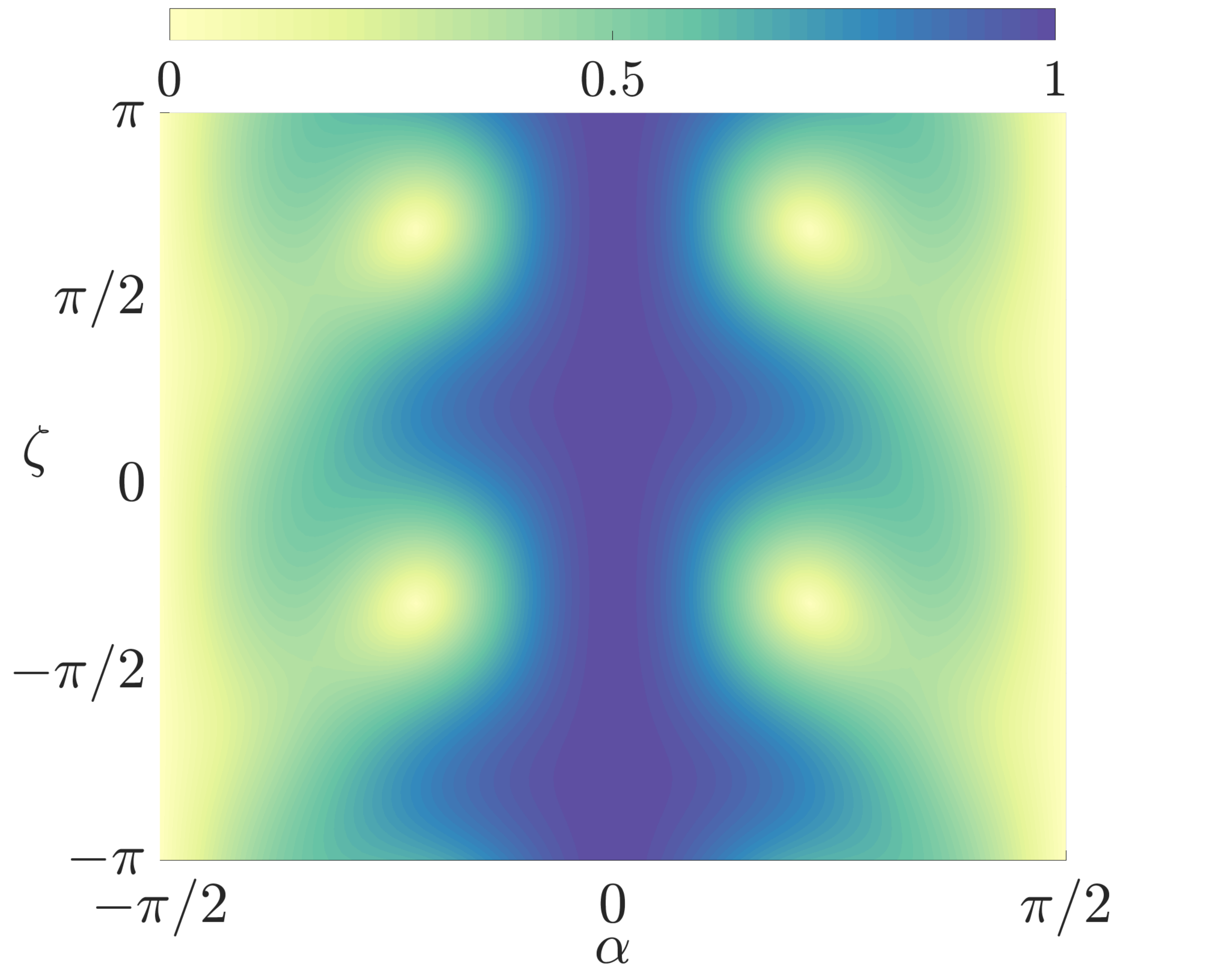}
    \put(0,88){(c)}
    \put(34,85){ \scalebox{.7}{$\sqrt{\dfrac{\Ceff^2 + \Deff^2}{\Ish^2 + \DIsh^2}}$} }
    \end{overpic}  
    \caption{Representations of the scaled effective chiral coefficients (a) $\Ceff/ \sqrt{\Ish^2 + \DIsh^2} \in [-1, 1]$, (b) $\Deff/ \sqrt{\Ish^2 + \DIsh^2} \in [-1, 1]$, (c) $\sqrt{(\Ceff^2 + \Deff^2)/(\Ish^2 + \DIsh^2)} \in [0, 1]$. We define these quantities in \eqref{eq: effective coefficients} and \eqref{eq: effective coefficients scaled}. Notably, the magnitude of each quantity is bounded above by one, so we may conclude that the effect of rapid rotation is to reduce the effective overall chirality of an object.}
    \label{fig:Effective_coeff}
\end{figure}

A helpful way to interpret the effective chirality parameters $\Ceff$ and $\Deff$, defined in \eqref{eq: effective coefficients}, is in terms of their relative sizes with respect to $\sqrt{\Ish^2 + \DIsh^2}$, which can be considered a measure of the overall chirality of the object. To study this, it is helpful to introduce the parameter $\zeta$, defined as the principle argument of the complex number
\begin{align}
\label{eq: alpha def}
\exp(i \zeta) = \dfrac{\Ish + i \DIsh}{|\Ish + i \DIsh|}.
\end{align}
Therefore, the introduction of $\zeta$ collapses the two-dimensional parameter space of $(\Ish, \DIsh)$ onto a single parameter via the complex unit circle. Then, utilising the relationship $\tan \angl = \ac$, where $\angl$ is the angle
between the rotational and helicoidal axis, we can re-write \eqref{eq: effective coefficients} as
\begin{subequations}
\label{eq: effective coefficients scaled}
\begin{align}
\dfrac{\Ceff}{|\Ish + i \DIsh|} &= \cos \angl \left(\cos^2 \angl \cos \zeta + \sin^2 \angl \sin \zeta \right), \\
\dfrac{\Deff}{|\Ish + i \DIsh|} &= \dfrac{\cos \angl}{2} \left(3\sin^2 \angl \cos \zeta + (2\cos^2 \angl- \sin^2 \angl) \sin \zeta \right),
\end{align}
\end{subequations}
which means we can illustrate the left-hand sides of \eqref{eq: effective coefficients scaled} in terms of just two parameters: $\angl$ and $\zeta$ (see Figure~\ref{fig:Effective_coeff}a,b). Through explicit calculation, it can also be shown that 
\begin{align}
\label{eq: chirality constraint}
\Ceff^2 + \Deff^2 \leqslant \Ish^2 + \DIsh^2,
\end{align}
which is illustrated in Figure~\ref{fig:Effective_coeff}c. Interpreting $\sqrt{\Ish^2 + \DIsh^2}$ as a measure of the overall chirality of the object, we can deduce that rotation never increases the overall effective chirality. In fact, in general, rotation reduces the overall chirality, only leaving the overall chirality unchanged for $\angl = 0$. While this reduction is a general property for the overall chirality, it is notable that \eqref{eq: effective coefficients} implies that rotation can cause specific individual chirality parameters to increase. That is, rotation can cause $|\Ceff| > |\Ish|$ or $|\Deff| > |\DIsh|$, but the constraint \eqref{eq: chirality constraint} means that these cannot occur at the same time. Since $\Ish$ and $\DIsh$ represent different aspects of chirality, we can interpret this as rotation allowing different \emph{aspects} of chirality to be over or underemphasised, even though rotation reduces the overall chirality of the object.

\section{Deriving the emergent translational dynamics}
\label{sec: emergent trans dyn}

Using the asymptotic expansions \eqref{eq: asy exp} in the transformed governing equations \eqref{eq: full gov eq translational trans}, we obtain the trivial leading-order (i.e. $\order{\omb}$) system
\begin{align}
\label{eq: full gov eq translational trans LO}
\pbyp{\Xvecpos_0}{\ts} = \bs{0},
\end{align}
which tells us that $\Xvecpos_0 = \Xvecpos_0(\tl)$.

At next order (i.e. $\order{1}$), we obtain the system
\begin{align}
\om \pbyp{\Xvecpos_1}{\ts} + \dbyd{\Xvecpos_0}{\tl} &= \vel + \Ypos_0 \e{3} - \shapecoeffb \left( \ehat{2} \ehat{3}^T - \ehat{3} \ehat{2}^T \right) \mat{E}^* \ehat{1} + \shapecoeffc \mat{E}^* \ehat{1} \notag \\
\label{eq: full gov eq translational trans FO}
&\quad + (\shapecoeffa - \shapecoeffc) (\ehat{1}^T\mat{E}^* \ehat{1}) \ehat{1},
\end{align}
with $2 \pi$-periodicity in $\ts$, recalling that $\vel = \velscala\ehat{1} + \velscalb\ehat{2} + \velscalc\ehat{3}$. The solvability condition that will give our emergent dynamics is obtained simply by averaging \eqref{eq: full gov eq translational trans FO} over $\ts \in (0, 2\pi)$. Performing this averaging and imposing periodicity in $\ts$, \eqref{eq: full gov eq translational trans FO} becomes
\begin{align}
\label{eq: full gov eq translational trans FO averaged}
\dbyd{\Xvecpos_0}{\tl} = \left<\vel + \Ypos_0 \e{3} - \shapecoeffb \left( \ehat{2} \ehat{3}^T - \ehat{3} \ehat{2}^T \right) \mat{E}^* \ehat{1} + \shapecoeffc \mat{E}^* \ehat{1} + (\shapecoeffa - \shapecoeffc) (\ehat{1}^T\mat{E}^* \ehat{1}) \ehat{1} \right>.
\end{align}
Some care needs to be taken in evaluating the right-hand side of \eqref{eq: full gov eq translational trans FO averaged}, since the swimmer-frame basis vectors $\ehat{i}$ are dependent on $\ts$ through their dependence on the Euler angles, with the explicit dependence given in \eqref{eq: lab to swimmer transformation}. Importantly, since the leading-order analysis is the same between Parts 1 and 2, and the first two terms on the right-hand side of \eqref{eq: full gov eq translational trans FO averaged} are present in Part 1, we can use our results of \S4.5 of Part 1 to immediately state that:
\begin{align}
\label{eq: V + shear}
    \left \langle \vel + \Ypos_0 \e{3} \right \rangle = \veleff \etilde{1}(\alp,\phic) + \Ypos_0 \e{3},
\end{align}
where
\begin{align}
\veleff := \dfrac{\velscala + \ac \velscalb}{\sqrt{1 + \ac^2}},
\end{align}
and $\etilde{1}(\alp,\phic)$ can be considered equivalent to the (hatted) basis vector $\ehat{1}$ in \eqref{eq: lab to swimmer transformation}, but with argument $(\theta,\phi)$ replaced by $(\alp,\phic)$. 

To calculate the remaining averages on the right-hand side of \eqref{eq: full gov eq translational trans FO averaged}, we start by writing them in terms of the laboratory basis, using the swimmer-to-laboratory transformation \eqref{eq: lab to swimmer transformation} and the definition of $\mat{E}^*$ \eqref{rosav}. This yields
\begin{subequations}
\label{eq: all E terms in lab frame}
\begin{align}
\left( \ehat{2} \ehat{3}^T - \ehat{3} \ehat{2}^T \right) \mat{E}^* \ehat{1} &= \dfrac{1}{2} \left( \left[s_{\theta}^2 c_{2\phi} \right] \e{1} + \left[ c_{\theta} s_{\theta} s_{\phi} \right] \e{2} + \left[ c_{\theta} s_{\theta} c_{\phi}  \right] \e{3}\right), \\
\mat{E}^* \ehat{1} &= -\dfrac{1}{2} \left( s_{\theta} c_{\phi} \e{2} - s_{\theta} s_{\phi} \e{3}\right), \\
(\ehat{1}^{T}\mat{E}^* \ehat{1})\ehat{1} &= -\dfrac{1}{2} \left( \left[c_{\theta} s_{\theta}^2 s_{2 \phi} \right] \e{1} + \left[s_{\theta}^3 s_{2 \phi} s_{\phi} \right] \e{2} - \left[s_{\theta}^3 s_{2 \phi} c_{\phi} \right] \e{3}\right),
\end{align}
\end{subequations}
where we have used shorthand notation with $c_{\theta}$, $s_{\theta}$, $c_{\phi}$, $s_{\phi}$, denoting $\cos \theta_0$, $\sin \theta_0$, $\cos \phi_0$, $\sin \phi_0$ etc. We can then calculate the averages of \eqref{eq: all E terms in lab frame} using the expressions \eqref{eq: mu def}, \eqref{eq: a 0 LO identities General a}--\eqref{eq: double angle relationships} we derived previously, to deduce that
\begin{subequations}
\label{eq: translational averages}
\begin{align}
\av{\left( \ehat{2} \ehat{3}^T - \ehat{3} \ehat{2}^T \right) \mat{E}^* \ehat{1}} &= \dfrac{2 - \ac^2}{4 \om^2} \left( \left[s_{\alp}^2 c_{2\phic} \right] \e{1} + \left[ c_{\alp} s_{\alp} s_{\phic} \right] \e{2} + \left[ c_{\alp} s_{\alp} c_{\phic}  \right] \e{3}\right), \notag\\
&= \dfrac{2 - \ac^2}{2 \om^2} \left( \etilde{2} \etilde{3}^T - \etilde{3} \etilde{2}^T \right) \mat{E}^* \etilde{1}, \\
\av{\mat{E}^* \ehat{1}} &= -\dfrac{1}{2 \om} \left( s_{\alp} c_{\phic} \e{2} - s_{\alp} s_{\phic} \e{3}\right) = \dfrac{\mat{E}^* \etilde{1}}{\om}, \\
\av{(\ehat{1}^{T}\mat{E}^* \ehat{1})\ehat{1}} &= -\dfrac{1}{4 \om^3} \Big[\left( 2 - 3 \ac^2\right)\big\{ \left[c_{\alp} s_{\alp}^2 s_{2 \phic} \right] \e{1} + \left[s_{\alp}^3 s_{2 \phic} s_{\phic} \right] \e{2} \notag \\
&\qquad 
- \left[s_{\alp}^3 s_{2 \phic} c_{\phic} \right] \e{3}\big\} + 2\ac^2 \left( s_{\alp} c_{\phic} \e{2} - s_{\alp} s_{\phic} \e{3}\right) \Big], \notag \\
&= \dfrac{2 - 3\ac^2}{2 \om^3} (\etilde{1}^{T}\mat{E}^* \etilde{1})\etilde{1} + \dfrac{\ac^2}{\om^3} \mat{E}^* \etilde{1},
\end{align}
\end{subequations}
where $\etilde{i} = \etilde{i}(\alp,\phic)$ and can be considered equivalent to their $\ehat{i}$ (hatted) versions in \eqref{eq: all E terms in lab frame} with arguments $(\theta_0,\phi_0)$ replaced by $(\alp,\phic)$.

\begin{figure}
\vspace{0.5em}
\begin{center}
\begin{overpic}[width=\textwidth]{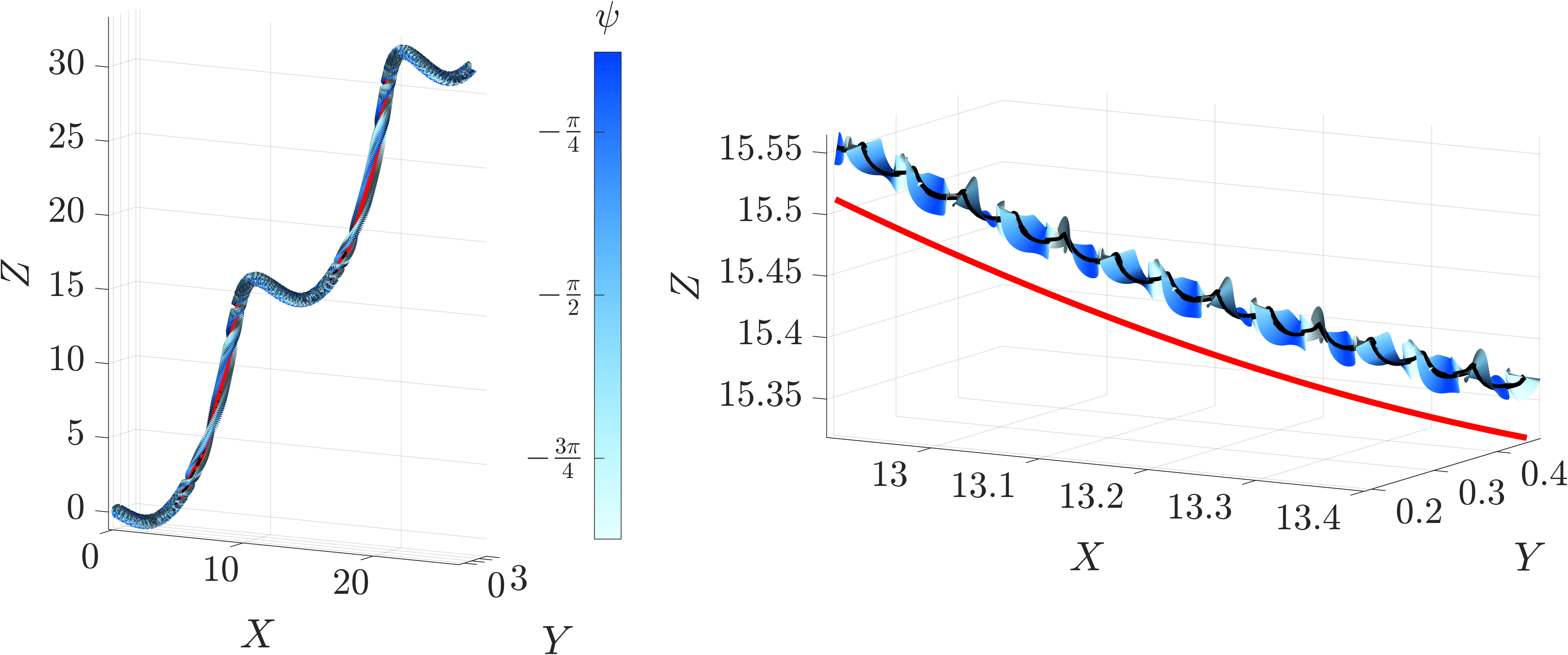}
    \put(-2,36){(a)}
    \put(45,36){(b)}
\end{overpic}
\caption{Illustration of the agreement between the full spinning translational dynamics and the emergent system we derive. (a) The predictions of the emergent dynamics are shown as a red curve, while the full dynamics are shown as a black line with attached ribbon, coloured according to the spin angle $\psi$ of the object. Differences between the dynamics are barely visible at the resolution of this plot. (b) A portion of the trajectory in (a), showing the small (expected) discrepancy between the full and emergent solutions. Here, we have taken $\omb = \oma = 100$, $\Breth=0.8$, $\Ish=-0.3$, $\DIsh=-0.5$, $\shapecoeffb=0.01$, $\shapecoeffc = 0.3$, $\shapecoeffa = -3$, $\velscala = 1$, $\velscalb = \velscalc = 0.5$.}
\label{fig:translational full vs averaged}
\end{center}
\end{figure}

\begin{figure}
\begin{center}
\includegraphics[width=\textwidth]{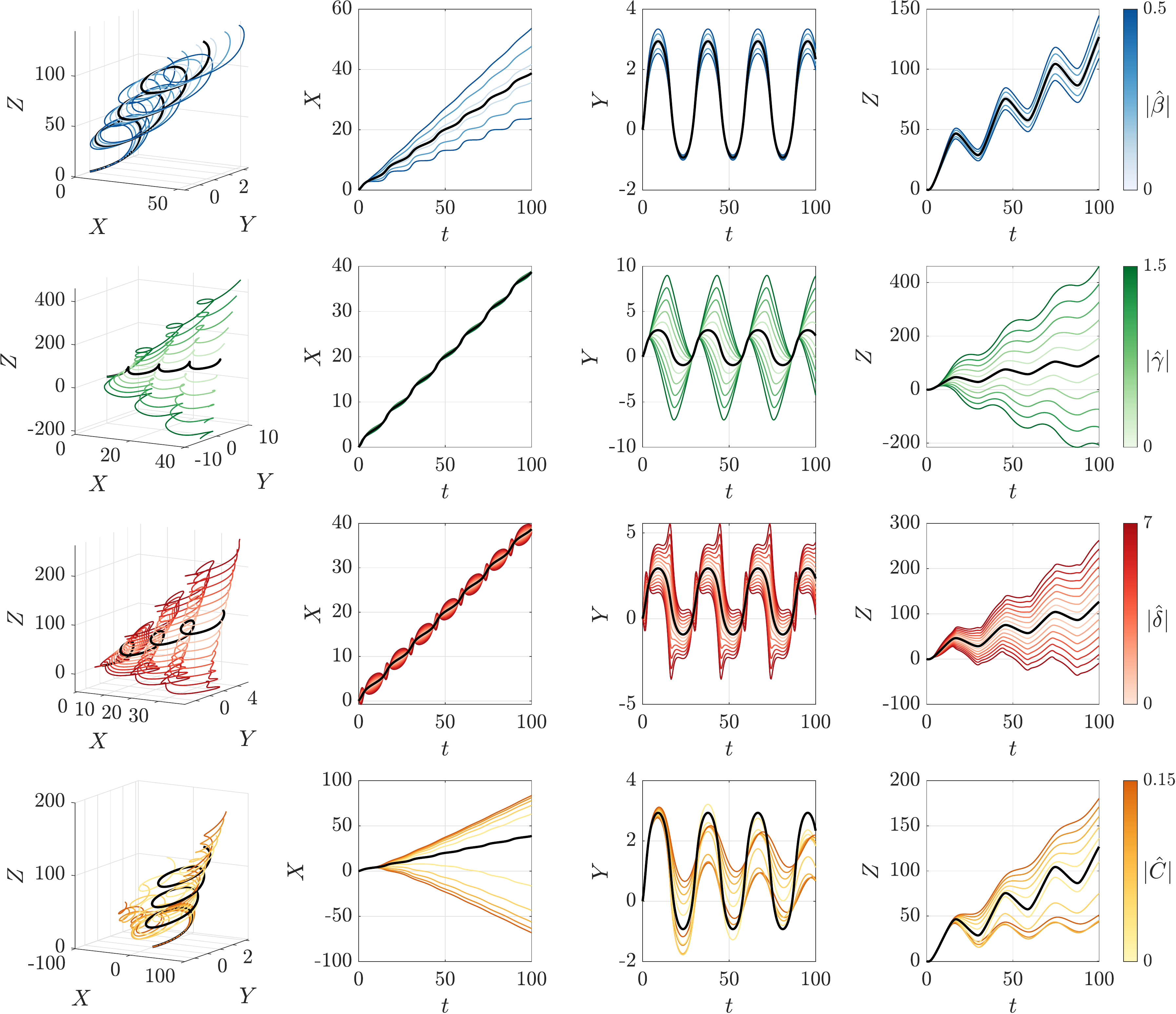}
\caption{Exploring the influence of the effective geometric parameters $\effshapecoeffb$, $\effshapecoeffc$, $\effshapecoeffa$, and $\Ceff$ on the emergent translational dynamics. In each row, we vary each effective parameter independently from $(\effshapecoeffb, \effshapecoeffc, \effshapecoeffa, \Ceff) = (0,0,0,0)$, highlighting the distinct role that each parameter plays in determining the emergent translational dynamics. In each column, we show three-dimensional trajectories and traces of laboratory-frame coordinates over time. Throughout, we use initial conditions $\vec{X} = 0$ and $(\theta,\phi,\psi) = (\pi/3,\pi/6,2\pi/3)$.}
\label{fig:translational without spinning}
\end{center}
\end{figure}

Finally, substituting \eqref{eq: V + shear} and \eqref{eq: translational averages} into \eqref{eq: full gov eq translational trans FO averaged}, we obtain our effective equation for the emergent translational dynamics:
\begin{align}
\label{eq: emergent translational equation}  
    \dbyd{\Xvecpos_0}{\tl} = \veleff \etilde{1} + \Ypos_0 \e{3} - \effshapecoeffb \left( \etilde{2} \etilde{3}^T - \etilde{3} \etilde{2}^T \right) \mat{E}^* \etilde{1} + \effshapecoeffc \mat{E}^* \etilde{1} + (\effshapecoeffa - \effshapecoeffc) (\etilde{1}^T\mat{E}^* \etilde{1}) \etilde{1},
\end{align} 
emphasising that $\etilde{i}$ are functions of the slow-time variables $\alp$ and $\phic$, and that we have defined the effective coefficients
\begin{align}
\label{eq: effective translation coefficients}
\effshapecoeffb = \dfrac{(2 - \ac^2) \shapecoeffb}{2(1 + \ac^2)}, \qquad
\effshapecoeffc = \dfrac{\shapecoeffc + \ac^2 \shapecoeffa}{(1 + \ac^2)^{3/2}}, \qquad
\effshapecoeffa = \dfrac{3 \ac^2 \shapecoeffc + (2 - \ac^2) \shapecoeffa}{2 (1 + \ac^2)^{3/2}},
\end{align}
and we illustrate these effective coefficients as functions of $\ac$ in Figure \ref{fig: bhat chat dhat}. Therefore, we see that the effective translational equation \eqref{eq: emergent translational equation} has the same functional form as the original equation \eqref{eq: full gov eq translational}, but with dependence on the fast-varying Euler angles switched to dependence on the slow-time functions we derived in \S\ref{Sec: emergent angular dynamics}, and modified coefficients \eqref{eq: effective translation coefficients} that systematically account for the effect of the fast spinning. Therefore, we can say that rapidly spinning chiral particles are translated as particles with an effective chiral shape, as quantified through the effective shape coefficients defined in \eqref{eq: effective translation coefficients}. The excellent agreement between the complex full translational dynamics and the emergent dynamics predicted by \eqref{eq: emergent translational equation} is illustrated on an example in Figure \ref{fig:translational full vs averaged}, and we explore further the effect of varying the intrinsic shape parameters in Figure \ref{fig:translational without spinning}.

Finally, we consider the limiting cases of $\ac \to 0$ and $|\ac| \to \infty$. In the limit of $\ac \to 0$, the effective coefficients are unchanged (i.e. $\effshapecoeffb \to \shapecoeffb$, $\effshapecoeffc \to \shapecoeffc$, $\effshapecoeffa \to \shapecoeffa$). That is, when the axis of rapid spinning coincides with the axis of helicoidal symmetry, the effective shape of the chiral swimmer is unchanged; the rapid rotation does not significantly impact the emergent translational dynamics. In contrast, in the limit of $|\ac| \to \infty$ the effective coefficients are changed, with $\effshapecoeffb \to -\shapecoeffb/2$ and  $\effshapecoeffc, \effshapecoeffa \to 0$. We recall that the results of \S\ref{Sec: emergent angular dynamics} state that the effective chirality coefficients also vanish in the same limit (i.e. $\Ceff, \Deff \to 0$ as $|\ac| \to \infty$), and that passive homochiral objects satisfy $\Ish = \DIsh = \shapecoeffc = \shapecoeffa = 0$ (see Figure \ref{fig:shapes}). Therefore, we may conclude that when rapid spinning occurs around an axis perpendicular to the axis of helicoidal symmetry, a general active helicoidal swimmer will behave as though it is a passive homochiral swimmer. This can be interpreted intuitively by noting that a rapidly rotating swimmer with rotation axis perpendicular to its helicoidal axis can be thought of as exhibiting a geometric rotational symmetry of $\pi$ around its rotation axis.

\section{Results and conclusions}
\label{sec: results}

 We investigated the emergent dynamics for a class of rapidly rotating active chiral particles with helicoidal symmetry, governed by the system \eqref{eq: full gov eq}--\eqref{eq: full gov eq translational}. We considered the problem where rotation is fast compared to external shear rate, with the rotation axis pointing in a general direction, fixed in the swimmer frame. Formally, we analysed the distinguished asymptotic limit $\omb, \oma \gg 1$ with $\ac = \oma/\omb = \order{1}$, noting that these quantities are related to the angle of the rotation axis from the symmetry axis $\angl$ via $\tan \angl = \ac$. From our analysis in \S \ref{Sec: emergent angular dynamics} and \S \ref{sec: emergent trans dyn} we found that, somewhat remarkably, the effect of rapid rotation can be readily incorporated into generalised Jeffery's equations with effective coefficients \eqref{eq: simplified reduced}, \eqref{eq: emergent translational equation}, as long as the emergent dynamics are defined in terms of appropriately transformed variables. This means that rapid rotation only modifies the emergent dynamics through changes in the effective shape parameters. That is, an active, rapidly spinning object exhibits the effective hydrodynamic shape of a (generally) differently shaped, non-spinning object. Moreover, our results characterise and quantify the specific hydrodynamic relationship between passive and rapidly spinning objects through explicit calculation of these effective parameters, each in terms of relevant original parameters and as a nonlinear function of $\ac = \tan \angl$.

Our analysis allows us to physically interpret the effect of rapid rotation on the emergent trajectories. As we discuss in more detail below, the effect of rotation off the helicoidal axis ($\oma$) is more important to the emergent dynamics than rotation on the helicoidal axis ($\omb$). Moreover, the broad effect of increasing $\angl$, the angle between the axes of rotation and symmetry, is to reduce the overall effective chirality of the effective hydrodynamic shape. Importantly however, moving the rotation axis away from the symmetry axis can over- and under-emphasise different aspects of chirality. This includes chiral aspects that do not significantly affect the dynamics of passive chiral objects in flow; as described below, our results show that rapid rotation can cause these aspects to become much more important for active helicoidal particles in flow.

For the spheroidal objects of Part 1, there is only one quantity, the Bretherton parameter $\Breth$, that characterises hydrodynamic interactions with the object. In contrast, as summarised in Table \ref{Table: Parameter types},  there are six shape parameters that specify the hydrodynamic interaction for a general helicoidal shape \citep{Ishimoto2020b}. Three of these arise in the orientational dynamics: $\Breth$, $\Ish$, and $\DIsh$, and the other three in the translational dynamics: $\shapecoeffb$, $\shapecoeffc$, and $\shapecoeffa$. As illustrated in Figure \ref{fig:shapes}, for a hydrodynamically achiral particle we have $\Ish = \DIsh = \shapecoeffb = 0$, and for a particle with hydrodynamic fore-aft symmetry (i.e. either homochiral or heterochiral) we have $\shapecoeffc = \shapecoeffa = 0$. We note that spheroids satisfy both of these constraints. Through our multiscale analysis, we have derived explicit forms for the effective versions of these parameters in \eqref{eq: effective B}, \eqref{eq: effective coefficients}, \eqref{eq: effective translation coefficients} (denoted with hats), which quantify and systematically account for the effects of rapid rotation in the system.

Notably, the presence of chirality and fore-aft asymmetry does not explicitly change the effective Bretherton parameter $\Beff$, defined in \eqref{eq: effective B}, from its equivalent expression in Part 1. However, the inclusion of these additional effects does impact upon the overall orientational dynamics of the emergent system \eqref{eq: simplified reduced}, since they introduce additional terms involving $\Ceff$ and $\Deff$ (defined in \eqref{eq: effective coefficients}) into the overall system. These two chirality parameters are the effective versions of $\Ish$ and $\DIsh$, respectively.  Notably, $\Ceff$ and $\Deff$ each depend on both $\Ish$ and $\DIsh$, and we show that $\Ceff^2 + \Deff^2 \leqslant \Ish^2 + \DIsh^2$ in \S \ref{sec: Summary orientational dynamics}. By interpreting $\Ish^2 + \DIsh^2$ as a measure of the overall hydrodynamic chirality of an object for its orientational dynamics, the effect of rotation is therefore to reduce the overall effective chirality of the object.

Interestingly however, rotation can cause $\Ceff > \Ish$ or $\Deff > \DIsh$ (though, from the constraint above, not both at the same time). Since $\Ish$ and $\DIsh$ reflect the moment of chirality along the helicoidal axis, this means that rapid rotation can enhance certain hydrodynamic aspects of chirality while reducing the overall hydrodynamic chirality of the object. Moreover, we note that the object `spin' $\psi(t)$ essentially decouples from the remaining variables in the full passive system (obtained by setting $\omb = \oma = 0$ in \eqref{eq: full gov eq}). Since $\DIsh$ only appears in \eqref{eq: psi eq}, the equation for $\psi$, this parameter is generally not important for the overall $(\theta,\phi)$ dynamics of the system, often the key observable dynamical outputs. However, our analytic results in \eqref{eq: effective coefficients} show that rotation can cause $\DIsh$ to significantly affect the effective coefficient $\Ceff$, which is important for the overall slow-time dynamics. This means that $\DIsh$ can be very important for the dynamics of rotating bodies, but unimportant for passive bodies. This effect could also explain why resistive force theory calculations give slightly smaller values for $\Ish$ than experimental estimates \citep{jing2020chirality, zottl2022asymmetric}. That is, theoretically calculated values of $\Ish$ for simple bacterium models can be fairly small, in contrast to $\DIsh$ (see e.g. estimates using resistive force theory calculations in Appendix \ref{sec: estimations}). However, since the effective Ishimoto parameter $\Ceff$ can be enhanced by $\DIsh$ in the presence of rotation, the (observed) effective Ishimoto parameter $\Ceff$ for spinning objects can be larger than for its passive equivalent $\Ish$.

An interesting implication of our results is that there are specific rotation rates and relationships between chirality parameters that cause both effective chirality parameters to vanish. Specifically, from \eqref{eq: effective coefficients} we see that a rotation axis satisfying $\tan^2 \angl = 2/3$ with chirality parameters satisfying $3 \Ish + 2 \DIsh = 0$ will result in $\Ceff = \Deff = 0$. This will result in the rotating object behaving hydrodynamically as an achiral object. While this requirement may be overly prescriptive to be observed in nature, it may be feasible to achieve for designed artificial swimmers. We note that this procedure is likely to involve a challenging optimisation process over the space of swimmer shapes, since the problem of finding a shape that satisfies specific coefficients is an inverse problem. This is in contrast to the less complex `forward' problem of calculating the shape coefficients from a given shape. It would be interesting in the future to solve the inverse problem of calculating object shapes that satisfy these constraints. Swimmers with these properties would behave as chiral objects when passive, and as achiral objects when rotating rapidly with $\tan^2 \angl = 2/3$. In addition, since a critical rotation ratio of $\tan^2 \angl = 2$ causes the effective Bretherton parameter $\Beff$ to vanish (so that the object is hydrodynamically equivalent to a sphere if it is achiral), it is not possible in general to prescribe a single critical rotation axis that causes $\Beff$, $\Ceff$ and $\Deff$ to vanish simultaneously.

The implications of our emergent translational dynamics \eqref{eq: emergent translational equation} have direct equivalence with the interpretation given above. This is because the effective shape parameters for translational dynamics in \eqref{eq: effective translation coefficients} are analogous to their orientational counterparts in \eqref{eq: effective B}, \eqref{eq: effective coefficients}. That is, the functional dependence on the rotation angle $\angl = \arctan \ac$ of the effective shape parameter $\effshapecoeffb$ in \eqref{eq: effective translation coefficients} is the same as that of the effective Bretherton parameter $\Beff$ in \eqref{eq: effective B}. Similarly, $\effshapecoeffc$ and  $\effshapecoeffa$ in \eqref{eq: effective translation coefficients} have the same functional dependence on $\angl$ as $\Ceff$ and $\Deff$, respectively, in \eqref{eq: effective coefficients}. Therefore, all of our conclusions above for $\Beff$, $\Ceff$, and $\Deff$ in the orientational dynamics also hold for $\effshapecoeffb$, $\effshapecoeffc$, and  $\effshapecoeffa$, respectively, in the translational dynamics. Perhaps interestingly, as noted above, $\shapecoeffb$ arises from chiral effects and $\shapecoeffc$, $\shapecoeffa$ can arise from a lack of fore-aft symmetry of the object. Therefore, the implications for $\Beff$, the effective Bretherton parameter for rotation (here for chiral particles, and in Part 1 for spheroidal particles), can be extended to the effective translation coefficient $\effshapecoeffb$. Similarly, the implications for the effective chirality coefficients for rotation $\Ceff$ and $\Deff$ can be extended to the effective translation coefficients $\effshapecoeffc$ and $\effshapecoeffa$. Therefore, by direct analogy with the results highlighted above and in Part 1, specific rotation rates and parameter dependencies can remove hydrodynamic chiral and fore-aft asymmetry effects in the effective translational dynamics.

\section{Discussion}
\label{sec: Discussion}

This study is the second in a two-part series, in which we have explored the emergent dynamics of three-dimensional, rapidly spinning, helicoidal objects in shear Stokes flow. In Part 2, we have explored the behaviours of completely general helicoidal objects, generalising our results from the spheroidal swimmer shape we imposed in Part 1. We have used the method of multiple scales for systems to systematically derive effective governing equations for the object dynamics. We have found that, when written in terms of appropriately transformed variables, the emergent equations are the generalised Jeffery's equations for passive chiral objects derived in \citet{Ishimoto2020b}, with appropriately modified hydrodynamic coefficients that account for the effects of rotation.

Our multiscale approach was vital in explicitly calculating these modified parameters. We used the method of multiple scales for systems (e.g. see pp.~127--128 of \citet{dalwadi2014flow} or p.~22 of \citet{dalwadi2018effect}) to systematically derive the appropriate emergent equations, which involved solving a three-dimensional nonlinear leading-order system, and a non self-adjoint problem at next order. The analytic derivation of the effective parameters allowed us to interrogate the general effect of rapid rotation on the emergent dynamics of helicoidal objects in shear flow. We showed that rotation along the helicoidal axis had little effect on the emergent dynamics. However, rotation off this axis had a more significant effect. Broadly, off-axis rotation reduces the overall magnitude of the effective parameters for both achiral and chiral objects. More specifically, the general effect of increasing off-axis rotation is to bring the effective aspect ratio of objects closer to unity through the reduction in magnitude of the effective Bretherton parameter $\Beff$. For chiral objects, the general effect of increasing off-axis rotation is to reduce the overall effect of chirality.

A curious aspect of our analysis is the nature of the equivalence between the effective equations we derive and the generalised Jeffery's equations for inert particles. Specifically, this equivalence is only evident when the slow-time variables that arise from our analysis are written in terms of suitably transformed variables. Although the appropriate definitions for these slow-time variables are related to the `average' position of the object, their specific choice is not immediately apparent when they first arise in the analysis. The choice we make in specifically using $\alp(t)$, $\muc(t)$, and $\phic(t)$ in \eqref{eq: mu def} only appears to be justified once we finally derive the emergent equations \eqref{eq: simplified reduced}. This is in contrast to recent applications of multiscale analysis to two-dimensional swimming problems \citep{Walker2022,Walker2022a}, where the equivalence between slow- and fast-time variables are more apparent from the start.

A natural question to ask is whether our results can be extended to consider several swimmers. In general, the consideration of multiple swimmers would be significantly more challenging, partly due to the difficulties in calculating explicit hydrodynamic tensors that account for the orientation of several swimmers simultaneously. A specific sublimit in which it may be possible to adapt our results is the limit of dilute suspensions, where swimmers are well separated and swimmer-swimmer interactions are rare. In this case, it may be possible to extend our results to estimate probability distributions for organism orientation as a function of local shear rate, though this remains a subject for future work. Additionally, it may be possible to generate effective equations by applying our methodology to point-particle models for the alignment of particles via hydrodynamic interactions \citep{katuri2022arrested}.

To conclude, over this two-part study we have investigated the behaviours of rapidly spinning, three-dimensional, helicoidal objects in shear flow. We have shown that the emergent orientational and translational dynamics can be described by the dynamics of passive, differently shaped objects in appropriately transformed variables. Moreover, we have calculated analytic representations of the effective parameters that encode the effective hydrodynamic shape of these objects. In other words, our systematic analysis has highlighted that the angular behaviours of such spinning objects can be described by generalisations of Jeffery's orbits for effective passive objects, so that this study serves to complement the works of \citet{Bretherton1962}, \citet{Brenner1964a} and \citet{Ishimoto2020b,Ishimoto2020a} by further broadening the scope of Jeffery's classical study of objects in slow flow \citep{Jeffery1922}.

\vspace{1em}

\textbf{Acknowledgements.} M.P.D. is supported by the UK Engineering and Physical Sciences Research Council [Grant No. EP/W032317/1]. C.M. is a JSPS Postdoctoral Fellow (P22023) and acknowledges support by the JSPS-KAKENHI Grant-in Aid for JSPS Fellows (Grant No. 22F22023). B.J.W. is supported by the Royal Commission for the Exhibition of 1851. K.I. acknowledges JSPS-KAKENHI for Young Researchers (Grant No. 18K13456), JSPS-KAKENHI for Transformative Research Areas (Grant No. 21H05309), JST, PRESTO, (Grant No. JPMJPR1921) and JST, FOREST (Grant No. JPMJFR212N). \\

\textbf{Declaration of interests.} The authors report no conflict of interest.\\

\textbf{Data accessibility.} Minimal computer code for exploring the dynamics, as well as the scripts used to generate the figures in this study are available at \url{https://github.com/Clementmoreau/spinningswimmers}.  

\appendix

\section{Deriving the equations of motion}
\label{sec: derivation eom}

\setcounter{equation}{0}
\renewcommand{\theequation}{\thesection\arabic{equation}}

In this Appendix, we derive the equations of motion for a self-propelled helicoidal swimmer in a simple shear, introduced in \S \ref{sec: Governing equations}.

\subsection{Kinematics}

We take the origin of the swimmer frame $\Xvecpos = \Xpos \e{1} + \Ypos \e{2} + \Zpos \e{3}$ to be the centre of hydrodynamic mobility of the swimmer. Therefore, $\Xvecpos$ lies on $\ehat{1}$ \citep{Kim2005}. To specify the angular dynamics, we introduce the 
Euler angles, for which we use the $xyx$ convention, with $\phi\in[0,2\pi)$, $\theta\in[0,\pi]$, and $\psi\in[0,2\pi)$, noting that we interpret $\phi$ and $\psi$ modulo $2\pi$. In terms of the swimmer-fixed and laboratory frames, the transformation between basis vectors is given by 
\begin{align}
    \eihatmat &= \left(\begin{array}{c|c|c}
    c_\theta & s_\phi s_\theta & - c_\phi s_\theta \\ \label{eq: lab to swimmer transformation}
    s_\psi s_\theta &  \hphantom{+}c_\phi c_\psi  - s_\phi c_\theta s_\psi  &  \hphantom{+}s_\phi c_\psi  + c_\phi c_\theta s_\psi \\
   c_\psi s_\theta &  - c_\phi s_\psi  -s_\phi c_\theta c_\psi  & 
   -s_\phi s_\psi  + c_\phi c_\theta c_\psi
   \end{array}\right) \eimat , 
\end{align}
as illustrated in Appendix A of Part 1, with $c_\theta$, $s_\theta$ denoting $\cos\theta$, $\sin\theta$, and similarly for other angles.

Further, the Euler angle transformation also gives the relation between the angular velocity of the swimmer frame in the presence of flow, denoted $\angvel^f$, and the time derivatives of the Euler angles via
\begin{equation}
    \angvel^{f} = \dot{\phi}\e{1} + \dot{\theta}\eprime{2} + \dot{\psi}\ehat{1} = \sum \hat{\Omega}_p^f \ehat{p},
\end{equation}
which simplifies to 
\begin{equation}\label{ang1} 
    \begin{pmatrix}
        \dot{\theta}\\
        \dot{\psi}\\
        \dot{\phi}\\
    \end{pmatrix} 
    = 
    \begin{pmatrix}
        0 &  c_\psi  & -s_\psi \\
        1 & -s_\psi c_\theta/s_\theta & -c_\psi c_\theta/s_\theta    \\
        0 &  \hphantom{+}s_\psi/ s_\theta & c_\psi/s_\theta          
    \end{pmatrix}
    \begin{pmatrix}
        \hat{\Omega}^{f}_1 \\
        \hat{\Omega}^{f}_2\\
        \hat{\Omega}^{f}_3
    \end{pmatrix}.
\end{equation}
Writing $ \vec{x} = x \e{1} + y\e{2} + z\e{3}$ for the position of a general point in the domain, we consider the shear flow
\begin{equation}
    \flowvel(\vec{x}) = y\e{3} = Y \e{3} + (y-Y)\e{3},
\end{equation}
where we have decomposed the flow into its contribution at the origin of the swimmer frame, defining $\vec{V}^* = Y\e{3}$, and a disturbance $(y-Y)\e{3}$ relative to this. The associated rate of strain and fluid angular velocity are given by 
\begin{equation}\label{rosav}
\mat{E}^* = \frac{1}{2} \left(\vec{\nabla}\flowvel + (\vec{\nabla} \flowvel)^T\right) = \frac{1}{2} \left(\e{2}\e{3}^T + \e{3}\e{2}^T\right), \quad
\angvel^*  = \frac{1}{2} \vec{\nabla} \wedge \flowvel = \frac{1}{2} \e{1}.
\end{equation}

\subsection{Mechanics}

The grand mobility tensor formulation of \citet{Kim2005}, with no external flow and viscosity non-dimensionalised to unity, gives the general relations
\begin{equation} \label{gm1} 
    \begin{pmatrix}
        -\vec{V}\\
        -\angvel\\
        \vec{S}^*
    \end{pmatrix}
    = 
    \gmtensor
    \begin{pmatrix}
        \vec{F}\\
        \vec{T}\\
        \vec{0}
    \end{pmatrix}.
 \end{equation} 
The block entries of the grand mobility tensor relate the force, $\vec{F}$, and torque, $\vec{T}$, generated by the self-propulsion mechanism to the velocity, angular velocity and stresslet of the particle in a quiescent field, which we denote by $\vel$, $\angvel$, and $\vec{S}^*$, respectively.
 
Furthermore, noting that $\vec{F}$ and $\vec{T}$ are assumed to be invariant on imposing the external shear flow, we have the analogous relation
\begin{equation} \label{gm2} 
    \begin{pmatrix}
        \vel^* - \vel^f\\
        \angvel^* - \angvel^f\\
        \vec{S}^f
    \end{pmatrix}
    =
    \gmtensor
    \begin{pmatrix}
        \vec{F}\\
        \vec{T}\\
        \vec{E}^*
    \end{pmatrix},
 \end{equation}
where we recall that $\bm V^* = Y\e{3}$ is the undisturbed velocity of the external flow at the origin of the swimmer frame, $\angvel^*$ and $\vec{E}^*$ are given by \eqref{rosav}, and $\vel^f$, $\angvel^f$, and $\vec{S}^f$ are the velocity, angular velocity, and stresslet of the particle in the shear flow, respectively.

Using \eqref{gm1} to eliminate $\vec{F}$ and $\vec{T}$ from \eqref{gm2}, we can rewrite the force $\vec{F}$ and torque $\vec{T}$ in terms of the swimming velocities, $\vel$ and $\angvel$ to obtain the translational velocity expressions
\begin{equation} \label{gv001}  
    \vel^f = \vel + Y\e{3} - \tilde{\vec{g}}\vec{E}^*, \quad (\tilde{\vec{g}}\vec{E}^*)_i = \tilde{g}_{ipq}E_{pq}^*, \quad \tilde{g}_{iqp} = \tilde{g}_{ipq},
\end{equation} 
and the rotational velocity expressions
\begin{equation} \label{gv2}   
    \angvel^{f} = \angvel + \angvel^* - \tilde{\vec{h}} \mat{E}^*, \quad (\tilde{\vec{h}} \mat{E}^*)_i = \tilde{h}_{ipq} E^*_{pq}, \quad \tilde{h}_{iqp} = \tilde{h}_{ipq}.
\end{equation} 
The expressions for $\tilde{\vec{g}}\vec{E}^*$ and $\tilde{\vec{h}}\vec{E}^*$ are derived in \citet{Ishimoto2020a, Ishimoto2020b} for helicoidal objects. Using these expressions, we deduce that
\begin{equation}
\label{transveldrift}
  \vel^f = \vel + \Ypos \e{3} - \shapecoeffb \left ( \ehat{2} \ehat{3}^T - \ehat{3} \ehat{2}^T \right ) \mat{E}^* \ehat{1} + \shapecoeffc \mat{E}^* \ehat{1} + (\shapecoeffa - \shapecoeffc) (\ehat{1}^T\mat{E}^* \ehat{1}) \ehat{1},
\end{equation}
where $\shapecoeffb$ is a shape parameter corresponding to chiral effects (which vanishes for an achiral particle), and $\shapecoeffc$, $\shapecoeffa$ are shape parameters corresponding to fore-aft asymmetry effects (which vanish for a particle with hydrodynamic fore-aft symmetry), and
\begin{equation}\label{angveldrift}
 \angvel^{f} = \angvel + \angvel^* -\Breth \left ( \ehat{2} \ehat{3}^T - \ehat{3} \ehat{2}^T \right ) \mat{E}^* \ehat{1} + \Ish \mat{E}^* \ehat{1} + (\DIsh - \Ish) (\ehat{1}^T\mat{E}^* \ehat{1}) \ehat{1},
\end{equation}
where $\Breth$ and $\Ish$ are the Bretherton and Ishimoto parameters, and $\DIsh$ is an additional shape parameter generated by the chirality of the object.

\begin{figure}
    \centering
    \vspace{1em}
    \begin{overpic}[width=9cm]{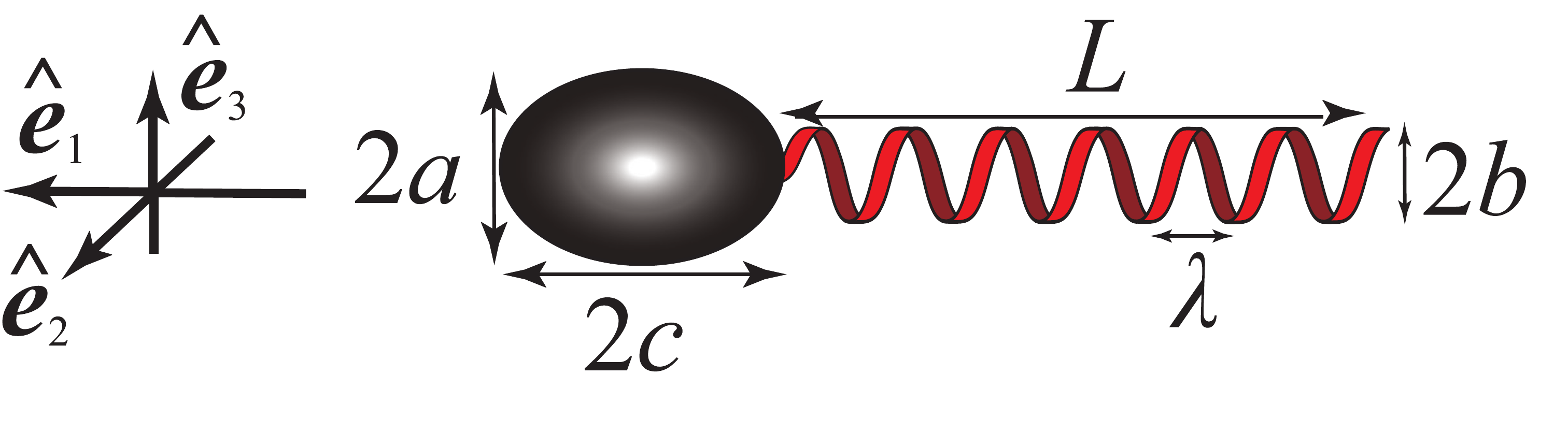}
    \put(-6,25){(a)}
    \end{overpic}
        \vspace{1em}
        
    \begin{overpic}[width=4.4cm]{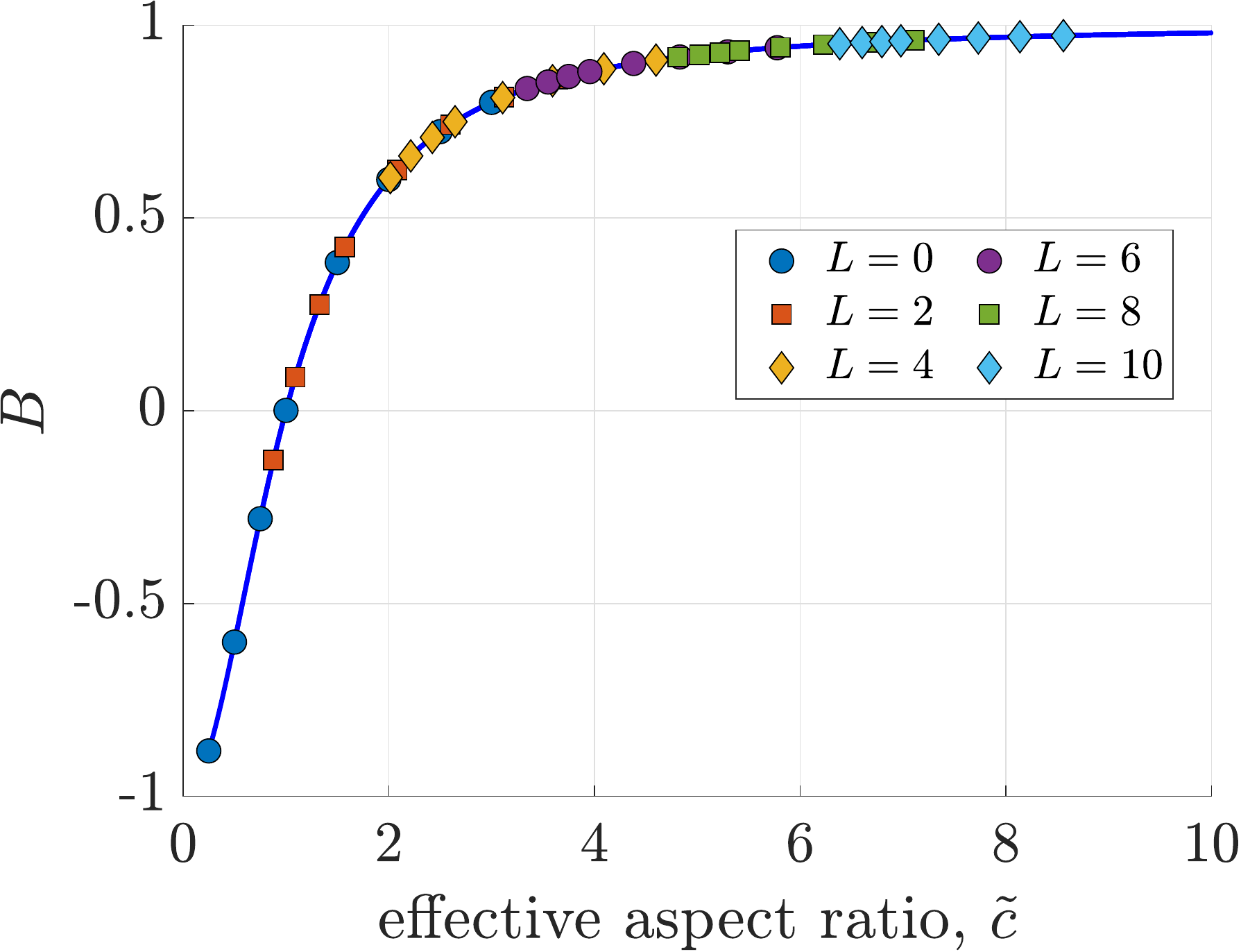}
    \put(0,81){(b)}
    \end{overpic}
    \begin{overpic}[width=4.4cm]{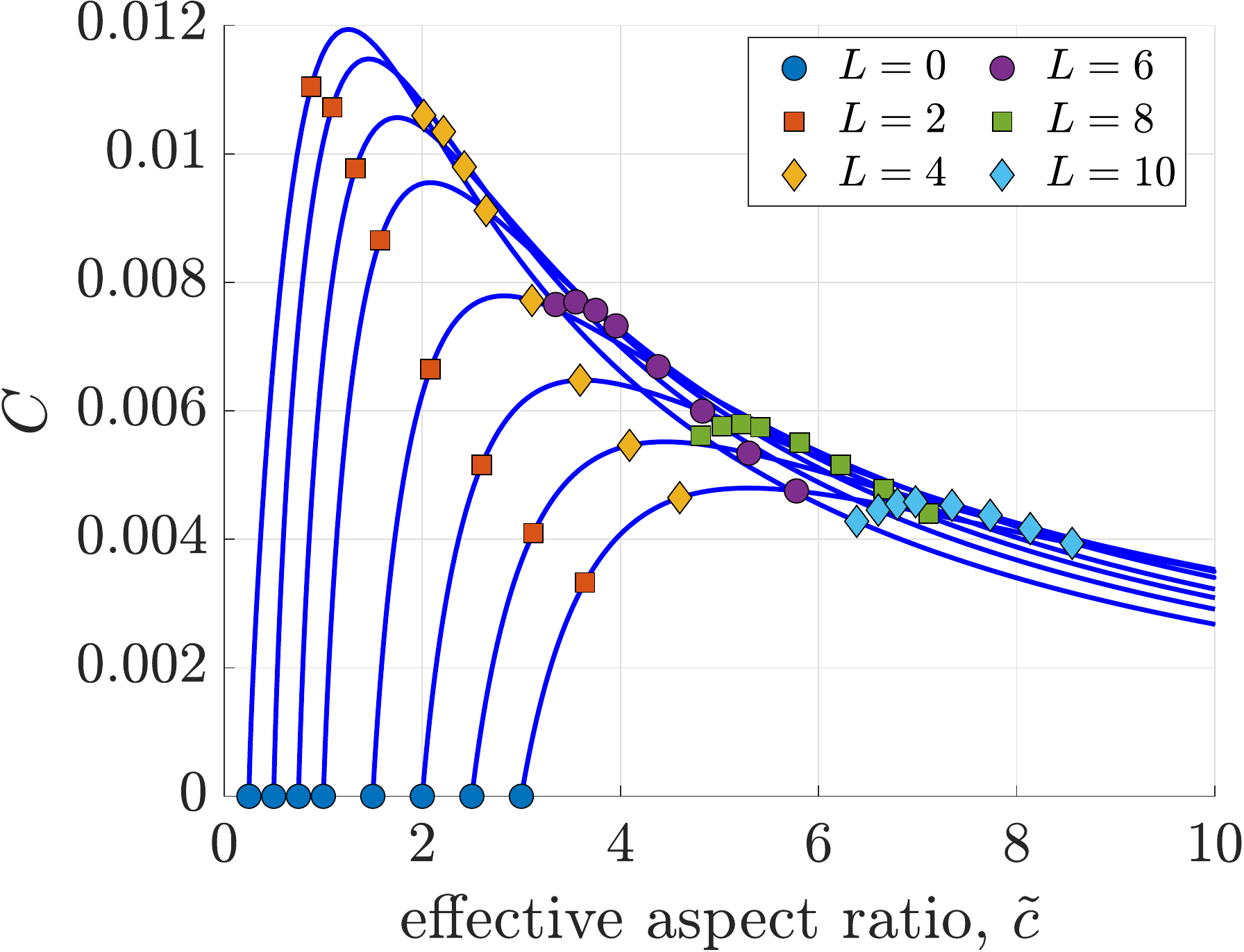}
    \put(0,81){(c)}
    \end{overpic}   
    \begin{overpic}[width=4.4cm]{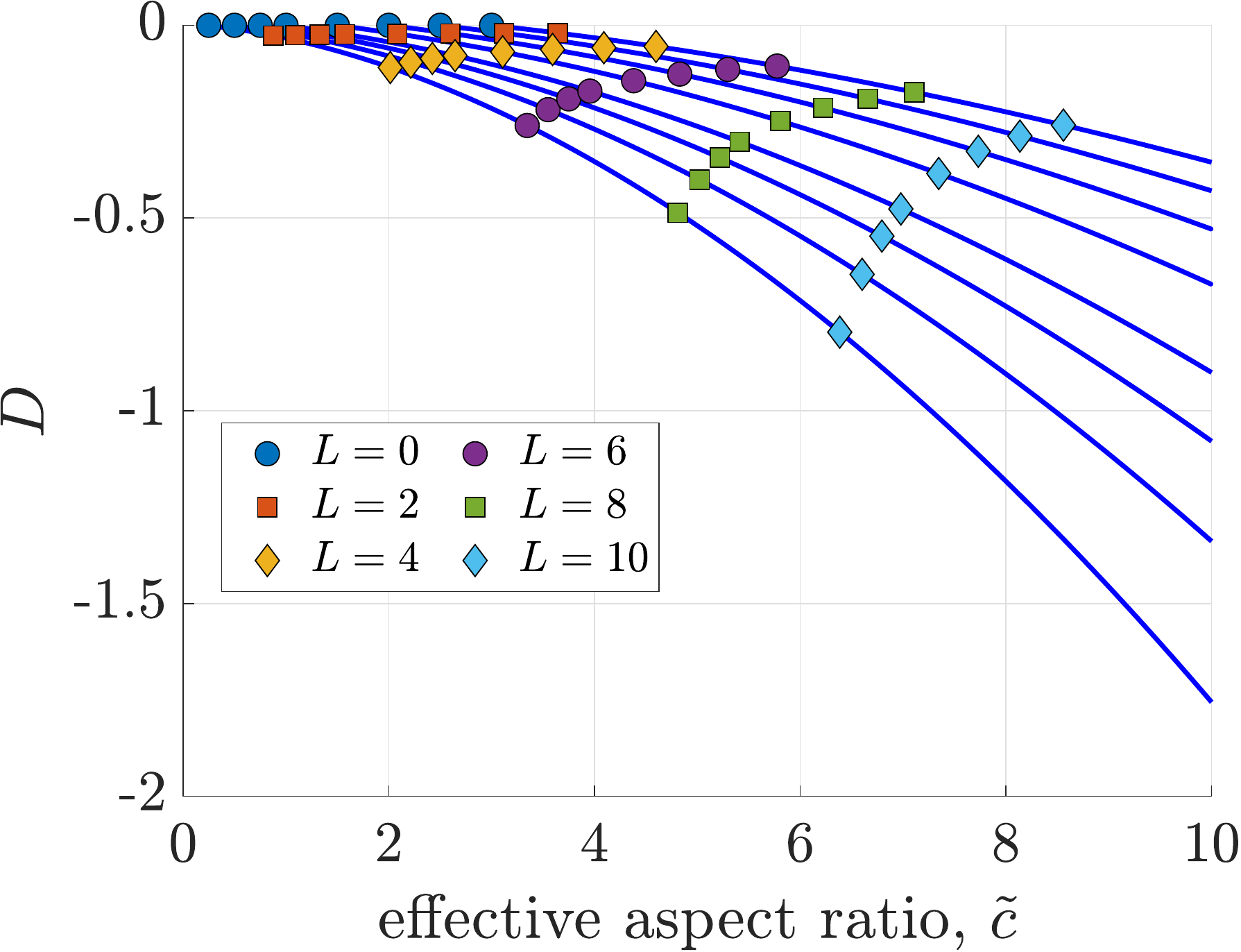}
    \put(0,81){(d)}
    \end{overpic}\\
    \vspace{2em}
    \begin{overpic}[width=4.4cm]{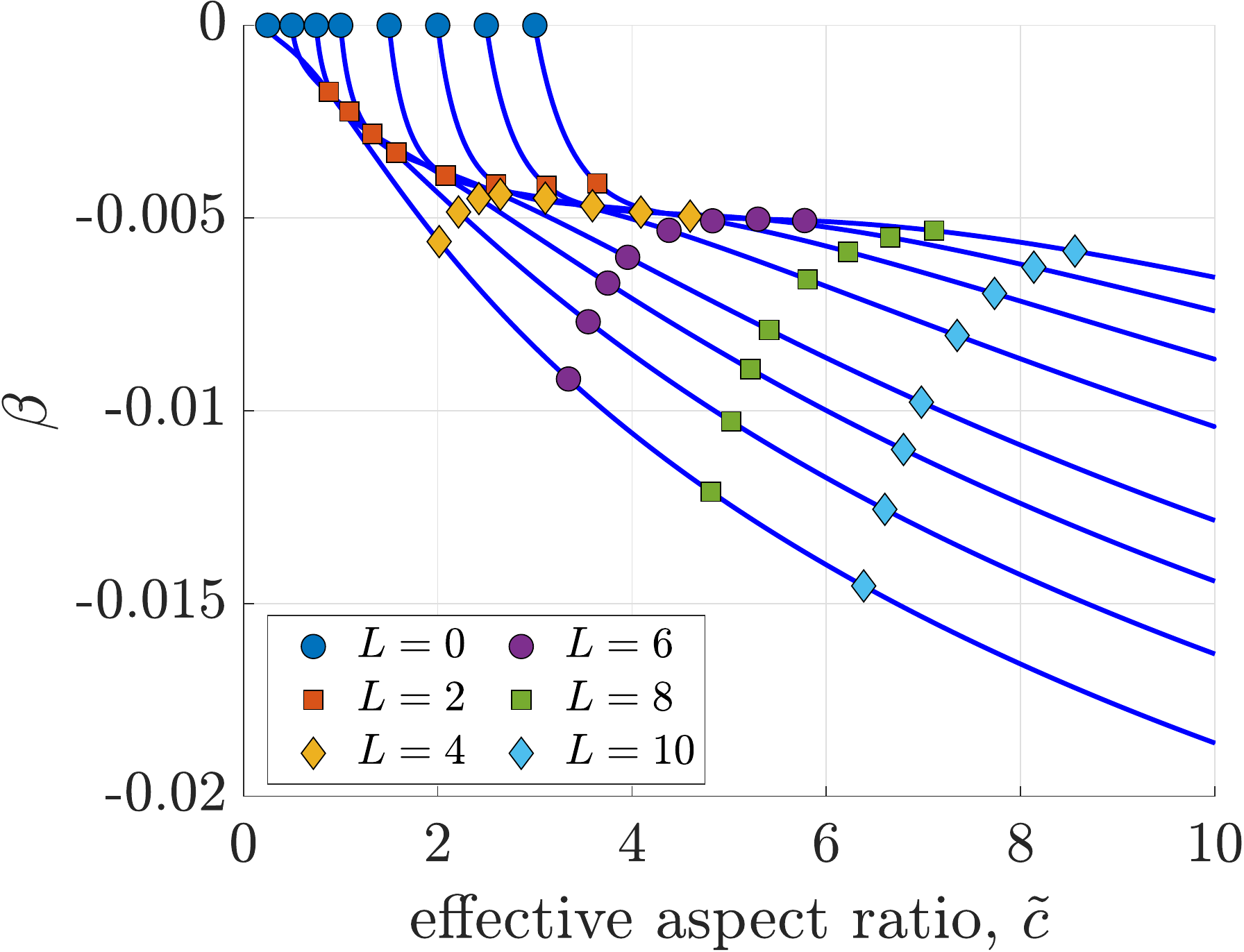}
    \put(0,82){(e)}
    \end{overpic}
    \begin{overpic}[width=4.4cm]{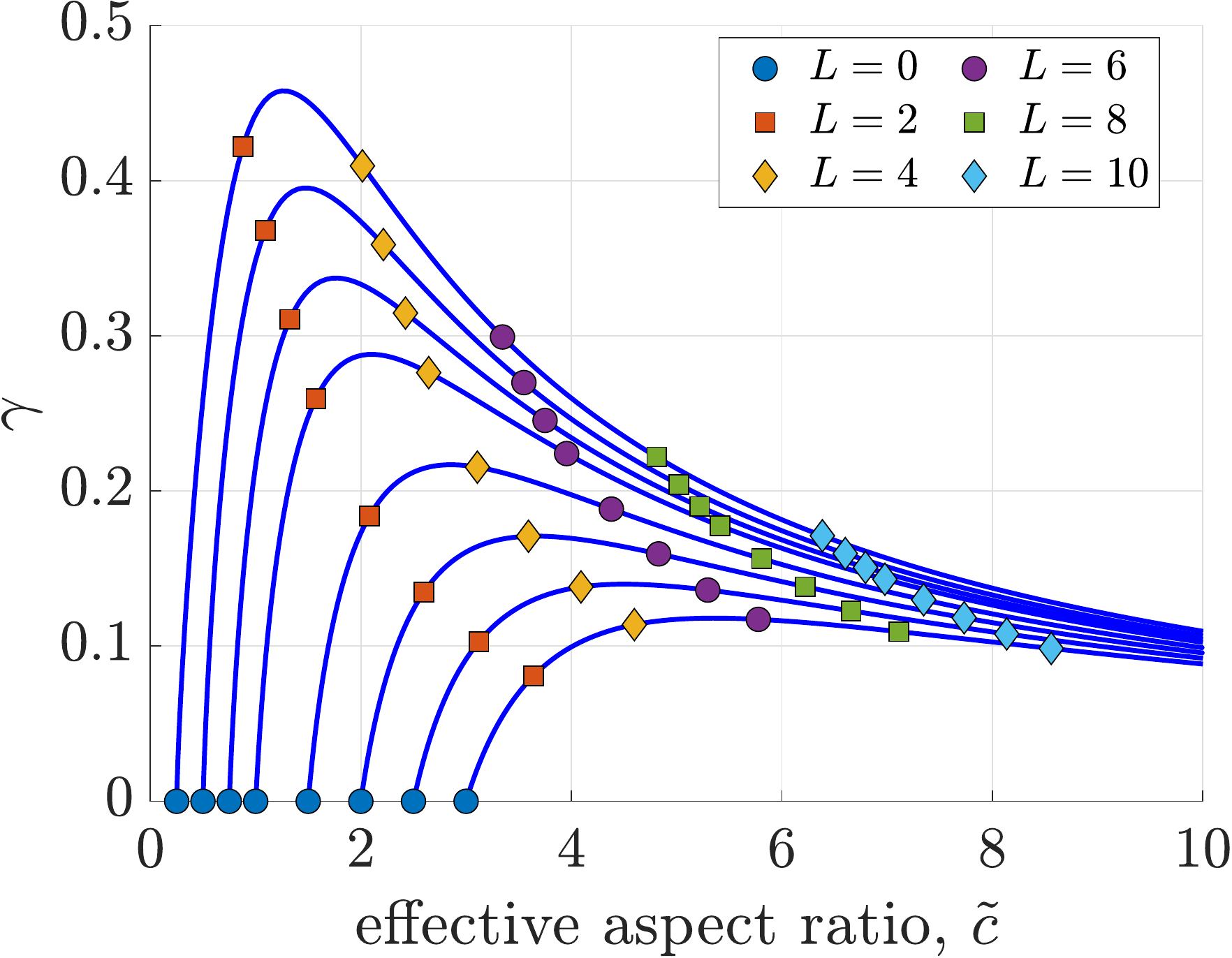}
    \put(0,82){(f)}
    \end{overpic}
    \begin{overpic}[width=4.4cm]{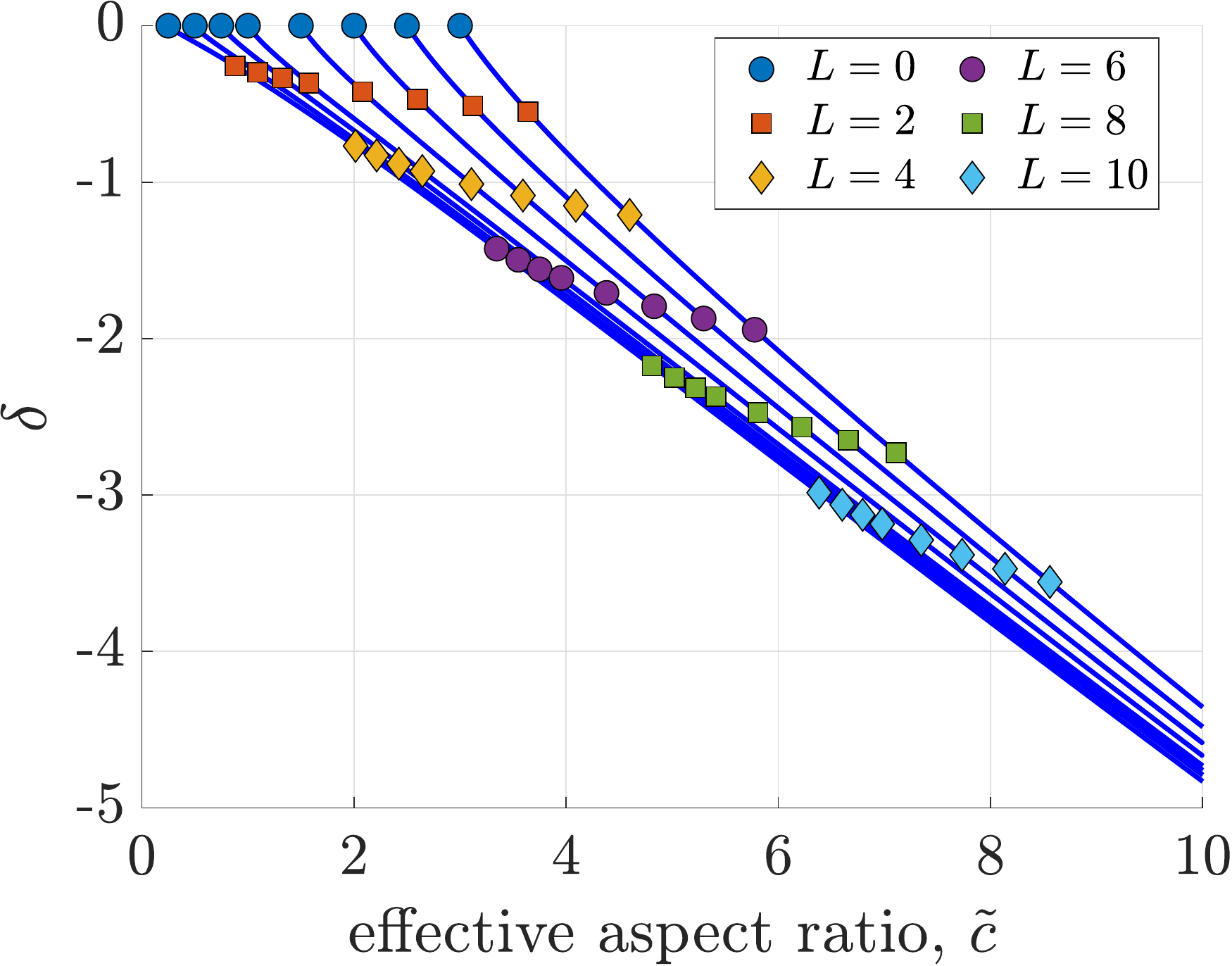}
    \put(0,82){(g)}    
    \end{overpic}    
    \caption{
    (a) Schematic of a model bacterium with a spheroidal cell body and a helical flagellum. The cell body is a rigid spheroid with semi-axes $c$, $a$, $a$, and the flagellum is a simple circular rigid helix with pitch $\lambda$, amplitude $b$ and length $L$. The axis of the helix is $\ehat{1}$, the director vector of the swimmer.     (b)-(g) The values of shape parameters, $\Breth$, $\Ish$, $\DIsh$, $\shapecoeffb$, $\shapecoeffc$, $\shapecoeffa$ for the model bacterium described in (a). These parameters are calculated from resistive force theory, using different cell body aspect ratios $c/a\in\{0.25, 0.5, 0.75, 1, 1.5, 2, 2.5, 3 \}$ and flagellar lengths $L$. The horizontal axis represents the effective aspect ratio, $\tilde{c}$, obtained from the values of $\Breth$. The remaining parameters are the same as those used in \citet{Ishimoto2020b} and $a=1$, $\lambda=2.5$ and $b = 0.25$. The lines represent different values of $c$. On each line, we use the symbols described in each legend to plot the values corresponding to $L=0,2, 4, 6, 8, 10$.}
    \label{fig:coeff_bact}
\end{figure}

Substituting \eqref{angveldrift} into \eqref{ang1} and using the frame transformation \eqref{eq: lab to swimmer transformation}
yields the angular dynamics given in  \eqref{eq: full gov eq}--\eqref{eq: g functions}. Recalling that $\textrm{d}\Xvecpos/\textrm{d}t=\vel^f$ and using \eqref{transveldrift} yields the translational dynamics given in \eqref{eq: full gov eq translational}.

\section{Estimation of shape parameters for a model bacterium}
\label{sec: estimations}

 In this Appendix, we estimate values of the shape parameters $\Breth$, $\Ish$, $\DIsh$, $\shapecoeffb$, $\shapecoeffc$, and $\shapecoeffa$ for a simple model bacterium used in a previous study \citep{Ishimoto2020b}. This simple model consists of a rigid spheroidal cell body (with semi-axes $c$, $a$, $a$) and helicoidal flagellum shown schematically in Figure \ref{fig:coeff_bact}(a). This left-handed simple helix has uniform circular cross-section, with radius $b$ and pitch $\lambda$. The flagellum axis and the semi-axis $c$ coincide with $\ehat{1}$, the axis of helicoidal symmetry. 
 
 We calculate the average values of hydrodynamic resistance around the $\ehat{1}$ axis using resistive force theory and the exact expression for a rigid spheroid. Analytic expressions for these quantities are provided in Appendix B of \citet{Ishimoto2020b}. We compute the shape parameters from their exact forms, represented by the components in the resistance matrix \citep{Ishimoto2020a}. The shape parameters $\Breth$, $\Ish$, $\DIsh$, $\shapecoeffb$, $\shapecoeffc$, $\shapecoeffa$ defined here correspond to $-\beta_2$, $\beta_3$, $\beta_1$, $-\alpha_2$, $\alpha_3$, $\alpha_1$, respectively, in \citep{Ishimoto2020a}. We plot these shape parameters in Figure \ref{fig:coeff_bact}(b)-(g) via the blue lines, with symbols denoting specific values of $L$.

 We vary the aspect ratio of the cell body $c/a$ and the flagellar length $L$ along the $\ehat{1}$ axis. Additionally, we fix $a=1$, $\lambda=2.5$ and $b = 0.25$. We use the cell body aspect ratios $c/a\in\{0.25, 0.5, 0.75, 1, 1.5, 2, 2.5, 3 \}$ and flagellar lengths $L$ from 0 to 100. The horizontal axis denotes the effective aspect ratio, $\tilde{c}$, obtained from the values of $B$ through the relationship $\tilde{c}=\sqrt{1+B}/\sqrt{1-B}$. For different values of $c$, we plot the values with $L=0,2, 4, 6, 8, 10$ using specific symbols.

\bibliographystyle{jfm_draft}
\bibliography{reference.bib,BJWMendeley.bib}

\end{document}